\documentclass[10pt,aps,onecolumn,superscriptaddress]{revtex4-2}

\usepackage[T1]{fontenc}
\usepackage[utf8]{inputenc}
\usepackage[english]{babel}
\usepackage{float}
\usepackage{amsfonts,amsbsy,amssymb,amsmath}
\usepackage{graphicx,float}
\usepackage{hyperref}
\hypersetup{
    colorlinks=true,
    linkcolor=blue,
    filecolor=magenta,      
    urlcolor=cyan,
    citecolor=blue
}

\graphicspath{{figures/}}

\begin{document}

\title{Bifurcation Study on a Degenerate Double van der Waals \\ Cirque Potential Energy Surface using Lagrangian Descriptors}

\author{Matthaios Katsanikas}
\email{matthaios.katsanikas@bristol.ac.uk}
\affiliation{School of Mathematics, University of Bristol, \\ Fry Building, Woodland Road, Bristol, BS8 1UG, United Kingdom.}

\author{Broncio Aguilar Sanjuan}
\email{broncio.aguilarsanjuan@bristol.ac.uk}
\affiliation{School of Mathematics, University of Bristol, \\ Fry Building, Woodland Road, Bristol, BS8 1UG, United Kingdom.}

\author{Francisco Gonz\'alez Montoya}
\email{fg16704@bristol.ac.uk}
\affiliation{School of Mathematics, University of Bristol, \\ Fry Building, Woodland Road, Bristol, BS8 1UG, United Kingdom.}

\author{V\'ictor J. Garc\'ia-Garrido}
\email{vjose.garcia@uah.es}
\affiliation{Departamento de F\'isica y Matem\'aticas, Universidad de Alcal\'a, \\ Madrid, 28871, Spain.}

\author{Stephen Wiggins}
\email{s.wiggins@bristol.ac.uk}
\affiliation{School of Mathematics, University of Bristol, \\ Fry Building, Woodland Road, Bristol, BS8 1UG, United Kingdom.}

\begin{abstract}

In this paper, we explore the dynamics of a Hamiltonian system after a double van der Waals potential energy surface degenerates  into a single well. The energy of the system is increased from the bottom of the  potential well up to the  dissociation energy, which occurs when the system becomes open. In particular, we study the bifurcations of the basic families of periodic orbits of this system as the energy increases using Lagrangian descriptors and Poincar\'e maps. We investigate the capability of Lagrangian descriptors to find periodic orbits of bifurcating families for the case of resonant, saddle-node and pitchfork bifurcations.  

\end{abstract}

\maketitle


\section{Introduction}

The study of bifurcations is an important topic in dynamical systems theory  with many applications in different areas such as chemical reaction dynamics, see \cite{inarrea2011bifurcations,founargiotakis1997bifurcation}. The changes in the phase space skeleton of the dynamics have important consequences for the global dynamics of such systems. In Hamiltonian systems, the KAM islands and the stable and unstable manifolds of normally hyperbolic invariant manifolds play an essential role in the dynamics. 
In recent years, different techniques have been developed to visualise the phase space like fast Lyapunov indicators (FLI) \cite{Lega2016}, mean exponential growth factor of nearby orbits (MEGNO) \cite{Cicotta2016}, the smaller (SALI) and the generalised (GALI) alignment indices \cite{Skokos2016}, delay time functions, scattering functions \cite{Gonzalez2012}, and Lagrangian descriptors \cite{madrid}. Lagrangian descriptors are a trajectory based diagnostic that can provide information about the stable and unstable manifolds of the unstable periodic orbits in a wide class of  systems. A remarkable property of these techniques is that they do not require that the trajectories intersect any particular predefined set of the phase space several times, a basic requirement for other tools such as Poincar\'e maps. These techniques are convenient tools for finding hyperbolic periodic orbits \cite{Guzzo2014,Demian2017,Lega2017}.

In this work, we analyse the changes in the phase space of a degenerate double van der Waals potential energy surface when the total energy is varied. The potential energy surface is constructed from the superposition of two identical wells such that their centres are very close. The van der Waals potential has been proposed to obtain an approximation to the escape time from a collision complex in an ultra-cold chemical reaction in \cite{Soley2018,Soley_thesis}. However, in the present work, we study the dynamics for energy $E<0$ when the trajectories are bounded. In one way, one can view the  degenerate double van der Waals potential is a perturbation of the integrable case with a very flat bottom in one direction. 

We use the method of Lagrangian descriptors to visualize the changes in the phase space structure after a bifurcation of periodic orbits and compare the results with those that are obtained using Poincar\'e sections. This paper aims to examine the capability of Lagrangian descriptors to detect hyperbolic periodic orbits close to their bifurcation. This is very important for providing an initial guess of the position of the hyperbolic periodic orbit because we can then use the guess to compute the new bifurcating families with classical methods of continuation of periodic orbits. In our case, we have two initially stable families of periodic orbits from the bottom of the well, which we call basic families. One family has periodic orbits that oscillate on the $x$ axis and the other family has periodic orbits  oscillating on the $y$ axis. We choose one of these families as an example, to present how the method of Lagrangian descriptors can give us valuable information on bifurcation problems. In particular, we study the capability of Lagrangian descriptors to detect the bifurcations of the basic family of the system, which has periodic orbits on the $y$-axis, that occur as the energy increases. In particular, these bifurcations are two resonant bifurcations and the other two symmetric saddle-node bifurcations connected with one subcritical bifurcation of the family of the well.

The paper is organized as follows. In Section \ref{sec:POS} we describe the construction of the potential energy surface that is the aim of this work. Section \ref{sec:methods} has a brief explanation of the Lagrangian descriptor method based on the Maupertuis action for Hamiltonian systems and the extraction of the stable and unstable manifolds of the periodic orbits using the Laplacian of the Lagrangian descriptor scalar field. In section \ref{sec:results}, we present the phase space analysis for the three cases of bifurcations of periodic orbits using  Lagrangian descriptors and comparing the results with Poincar\'e maps. Furthermore, we present the stability diagrams of the families of periodic orbits for the cases where bifurcations occur. Finally, we present the conclusions and final remarks in Section \ref{sec:conclutions}.

\section{The Potential Energy Surface}
\label{sec:POS}

\begin{figure}[htbp]
    A)\includegraphics[width=0.38\textwidth]{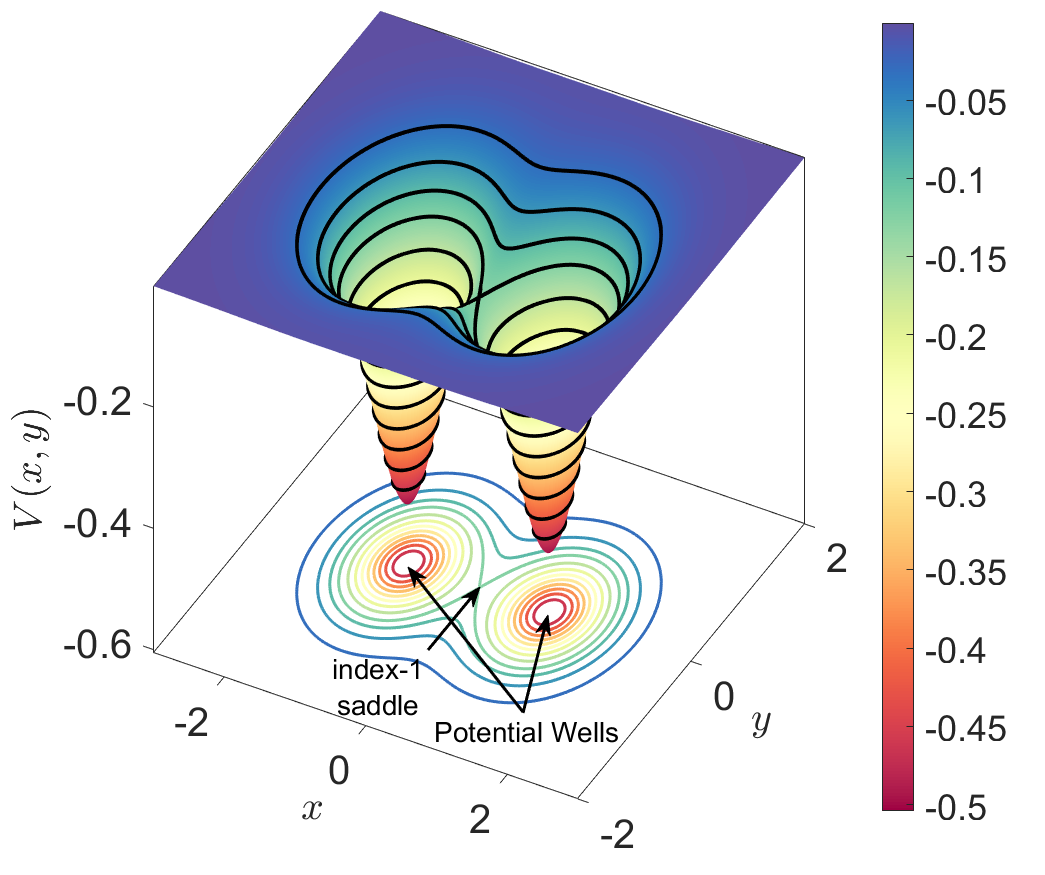}
	B)\includegraphics[width=0.33\textwidth]{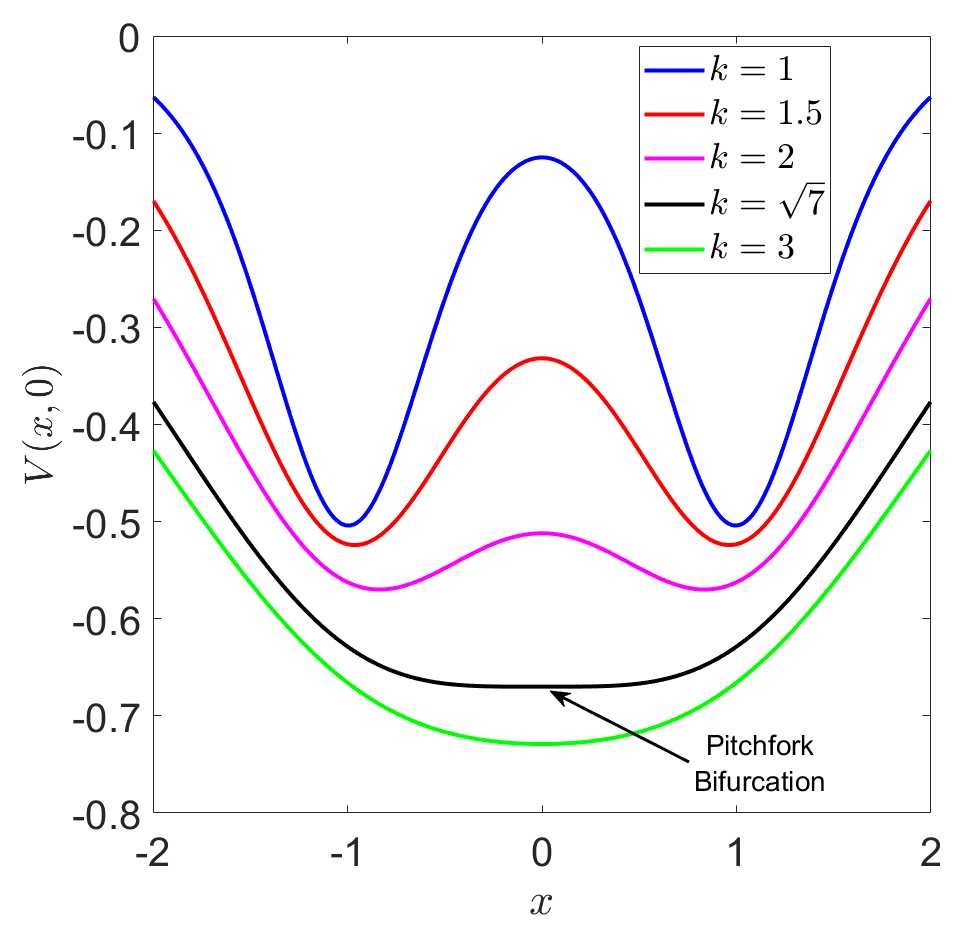}
	C)\includegraphics[width=0.38\textwidth]{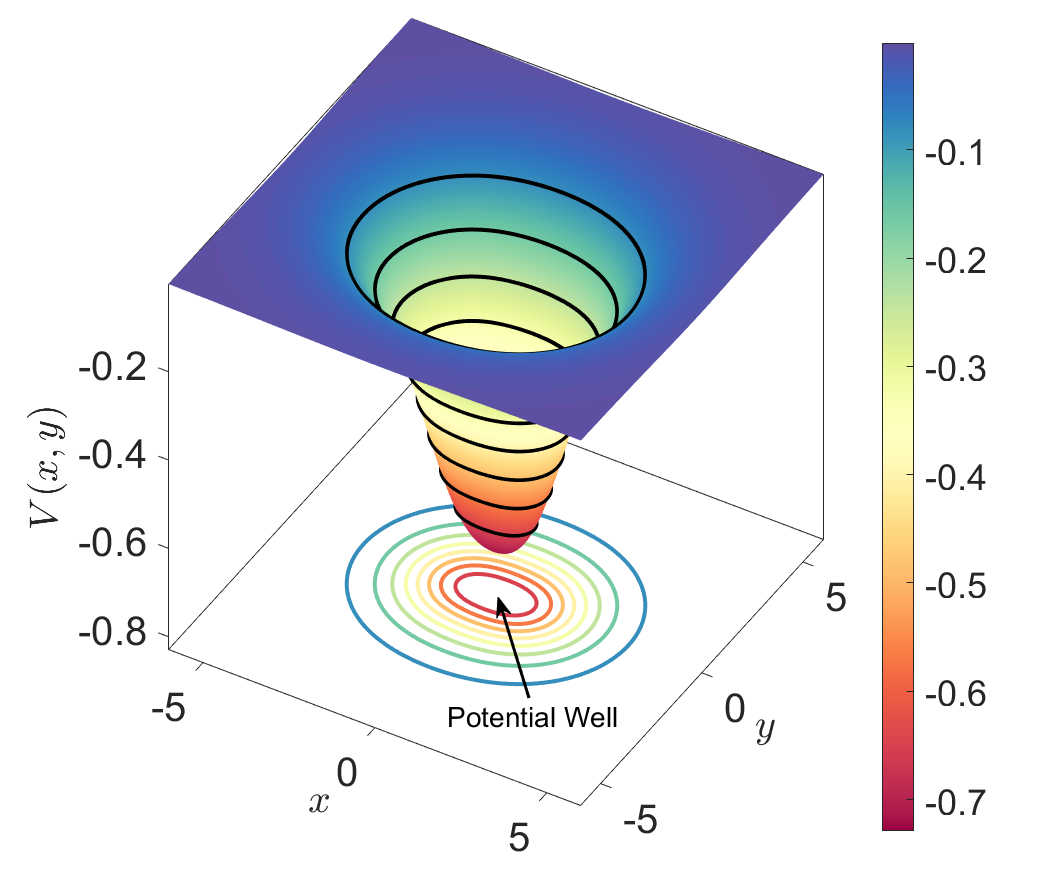}
	\caption{
	Degeneration of a double van der Waals (A) into a single well potential energy surface (C) with a non-hyperbolic fixed point at the origin, as a result of a \emph{pitchfork bifurcation} of the potential (B) along the $x$-axis when the parameter controlling the well-width, $k$, is varied in Equation \eqref{eq:double_vdw}, with critical value $k = d \sqrt{7}$, when $d$ fixed.
	Model parameters: (A) $W_0 = 1/2$, $k = 1$, $d = 1$; and (C) $W_0 = 1/2$, $k = \sqrt{7}$, $d = 1$.
	}
	\label{fig:double_vdw_PES}
\end{figure}

In this section we introduce and describe the construction of the potential energy surface (PES), $V(x, y)$, of the Hamiltonian describing the dynamics of the system in this work. The Hamiltonian is defined as the classical sum of kinetic energy plus potential energy:
\begin{equation}
H(x,y,p_x,p_y) = \dfrac{p_x^2}{2 m_1} + \dfrac{p_y^2}{2 m_2} + V(x,y)
\label{eq:hamiltonian}
\end{equation}
where the PES is given by:
\begin{equation}
V(x, y) =  \dfrac{- W_0 \, k^6}{\left(\left(x - d\right)^2 + y^2 + k^2\right)^3} + \dfrac{- W_0 \, k^6}{\left(\left(x + d\right)^2 + y^2 + k^2\right)^3}
\label{eq:double_vdw}
\end{equation}
We study the dynamics in the case where the model parameters take the values $W_0 = 1/2$, $d = 1$ and $k = \sqrt{7}$. This corresponds to the configuration where the PES has a minimum, i.e. a well, at the origin which is a degenerate equilibrium point, as discussed in Appendix \ref{appx:A}. The topography of the PES is depicted in Fig. \ref{fig:double_vdw_PES}C. The PES in Eq. \eqref{eq:double_vdw} is in general a two-well PES obtained from the superposition of two identical van der Waals potential wells \cite{Soley2018, Soley_thesis}, see Fig. \ref{fig:double_vdw_PES}A. The parameters $d$, $W_0$, and $k$ control, respectively the separation distance, the depth, and the width of both wells. But when the width of the wells increases simultaneously they can eventually merge forming a single well, see Fig. \ref{fig:double_vdw_PES}B. In the process, the PES undergoes a pitchfork bifurcation due to a change in stability of the saddle point as the wells merge, see Appendix \ref{appx:A} for details. 

In this work we study the dynamics of this system in the energy regime that goes from the energy of the PES at the origin (the well), $E = V(0,0)=-(7/8)^3$, up to zero, which is the dissociation energy. Due to the symmetries of $V(x,y)$ with respect to the axes, there are 2 families of periodic orbits associated with the minimum of the degenerate well. One family has  periodic orbits that oscillate along the $x$-axis and the other has periodic orbits that oscillate along the $y$-axis. As an example to illustrate the capabilities of the Lagrangian descriptor method, we only study the family with the  periodic  orbits that oscillate in the $y$ direction.

% parabolic direction x
% elliptical direction y

% In this work, we analyse the dynamics of this degenerated potential which is obtained from Eq.\eqref{eq:double_vdw} when $W_0 = 1/2$, $d = 1$ and $k = \sqrt{7}$. Here, we constructed the PES in Eq. \eqref{eq:double_vdw} inspired by the literature in chemical reaction dynamics. In such context, this type of potential is usually referred as \emph{cirque potential} \cite{Quapp2007, Kraka2010}. Also, in connection with that context, the single well potential in Eq.\eqref{eq:heller_pes} is equivalent to the one introduced in Ref.\cite{Soley2018}, Eq.(7) introduced to model long-range weak attractive interaction between neutral molecules/atoms, also known as \emph{van der Waals forces}. It is out of the scope of this paper to make a further connection with this context. However, for a further introduction to the relevance of cirque potentials like the one here and their dynamics, we direct the reader to Refs. \cite{Carpenter2017, Carpenter2018, GonzalezMontoya2020}.

%%%%%%%%%%%%%%%%%%%%%%%%%%%%%%%%%%%%%%%%%%%%%%%
% METHODS
%%%%%%%%%%%%%%%%%%%%%%%%%%%%%%%%%%%%%%%%%%%%%%%%

\section{The Action-based Lagrangian Descriptor}
\label{sec:methods}

In this section, we describe the method of Lagrangian Descriptors (LDs) that we employ to visualize structural changes in the phase space that characterize the underlying dynamics of the Hamiltonian system. Poincar\'e maps were also computed to complement and compare the information offered by LDs.

The method of LDs is a trajectory diagnostic that relies on the evaluation of a scalar field over sets of initial conditions to highlight the phase space location and geometry of invariant manifolds associated with periodic orbits and equilibria. This scalar field, given by the LD function, is often defined as a mapping of the initial condition of a trajectory to an ``arc length'' integral under some metric along the same trajectory \cite{mancho2013lagrangian}. Thus, when trajectories are resolved forwards (backwards) in time from a chosen set of initial conditions, stable (unstable) invariant structures intersecting with this set become distinguishable from abrupt changes in the LD output image. Unlike traditional methods to study phase structure, such as Poincar\'e maps, recurrence of trajectories is not required; hence, the LD map can always be computed \cite{ldbook2020}. The flexibility of the method has also allowed for adaptations and extensions to be able to study systems in different settings: autonomous and non-autonomous Hamiltonian \cite{mancho2013lagrangian}, discrete \cite{Lopesino2015, lopesino2017, Garcia-Garrido2018}, and stochastic \cite{Iniesta2016}. As a result, the method has found applications in fields such as fluids, for identification of coherent structures driving mixing and transport \cite{mendoza2010}; and chemistry, to study the phase space structures driving the dynamics of a reaction \cite{naik2019a,katsanikas2020a,katsanikas2020detection,katsanikas2020phase,crossley2021poincare,naik2020, krajnak2019, GG2020a, GG2020b, ,agaoglou2020phase,agaoglou2019}. For an extensive review on the theory of LDs and computation, see the online book \cite{ldbook2020}.

In this study, we select the action-based LD as the tool to study the dynamics in phase space like in \cite{Montoya2020}. The action-based LD function, $M_{S_0}$, is given as the sum of the forwards and backwards action-based LD functions, $M_{S_0}^+$ and $M_{S_0}^-$, here, expressed as:
\begin{equation}
    M_{S_0}^+ = \int_{t_0}^{t_0 + \tau_+ } \left[ p_x \frac{dx}{dt} + p_y \frac{dy}{dt} \right] dt \quad,\quad
    M_{S_0}^- = \int_{t_0 - \tau_- }^{t_0} \left[ p_x \frac{dx}{dt} + p_y \frac{dy}{dt} \right] dt
    \label{eq:action_ld}
\end{equation}
where $\tau_+, \tau_- > 0$ are respectively the forwards and backwards integration times of trajectories. As required, integrals are functions of the initial condition $(x_0, y_0, p_{x0}, p_{y0})$ at initial time $t_0$. Thus, the total Lagrangian descriptor is:
\begin{equation}
    M_{S_0} = M_{S_0}^{+} + M_{S_0}^{-} \;.
\end{equation}
As shown in \cite{GonzalezMontoya2020}, stable (unstable) invariant structures are accentuated by abrupt changes in the mapping of $M_{S_0}^{+(-)}$. The integrals $M_{S_0}^{+(-)}$ are based on the Maupertuis' action, \emph{a.k.a} the abbreviated action, $S_0$, which is related to Hamilton's action, $S$, from the least-action principle. Maupertuis' action is originally formulated as a line integral along a path with end points, $\mathbf{q}_A$ and $\mathbf{q}_B$, as:
\begin{equation}
S_0 = \int_{\mathbf{q}_A}^{\mathbf{q}_B} \mathbf{p} \cdot d\mathbf{q} \,.
\end{equation}
But, expression  Eq. \eqref{eq:action_ld}, clearly equivalent, facilitate the computation of $M_{S_0}^{+(-)}$ via evaluation of Hamilton's vector field along trajectories, which are usually numerically computed within a given time interval anyway. In contrast to LD versions based on the arc length of a trajectory under some space norm, i.e., $L_p$-norm, the action-based LD offers a dynamic interpretation, not only geometric. The integrals \eqref{eq:action_ld} are equivalent to time integrals of the kinetic energy of the system:
\begin{equation}
\int_{t_A}^{t_B} 2 \sqrt{E - V(\mathbf{q}(t))} \, dt \,.
\end{equation} 
Thus, high (low) values of $M_{S_0}$ are indicative of initial conditions whose trajectories are 'fast' (or 'slow'), provided that time intervals are identical, and the energy of the system, $E$, is constant \cite{GonzalezMontoya2020}.

Here, all LD calculations were performed using \texttt{ldds}, a Python package for computing and visualizing Lagrangian Descriptors in dynamical systems. See the GitHub repository for details in the implementation and easy-to-follow tutorials explaining the setup for computation of LDs \cite{LDDS}. Although the method of LDs can emphasize structures in phase space, we also applied a Laplacian-based filter to all LD maps to isolate the invariant manifold curves bounding POs. We computed the Laplacian values of LD maps using the \texttt{scipy.ndimage.laplace} Python routine \cite{scipy_laplace}, then took the square of these values and filtered out all points in the visualisation grid with values under a threshold. Threshold values were manually selected as needed for improved visualisation of the shape and location of the invariant curves. The resulting curves corresponding to stable and unstable manifolds were coloured in blue and red, respectively, for their distinction. 

%%%%%%%%%%%%%%%%%%%%%%%%%%%%%%%%%%%%%%%%%%%%%%%%%%%%
%
% Results and Discussion
%
%%%%%%%%%%%%%%%%%%%%%%%%%%%%%%%%%%%%%%%%%%%%%%%%%%%%

\section{Results and Discussion}
\label{sec:results}

In this section, we study the bifurcations of the basic family of periodic orbits in the potential well system using Lagrangian descriptors. To check this ability of LDs to detect bifurcations of periodic orbits, we compute three cases of bifurcations using the well-established method of the H\'enon stability diagrams. This technique is based on a standard continuation method to follow the stability of a family of periodic orbits versus a parameter (energy) of the system and the classical method of Newton-Raphson to find the location of the periodic orbits (see for example \cite{skokos2002orbital,contopoulos2004order}). Then, we study the capability of the method of the Lagrangian descriptors to detect these bifurcations of periodic orbits and comparing it with the method of Poincar\'e map. Due to the  symmetry under reflections with respect to the $x$-axis of the system, we consider the Poincar\'e surface of section  $y=0$ with $p_y>0$ as a set of initial conditions.

The first case of the bifurcations is a resonant bifurcation of periodic orbits with period 2 of the basic family of the system from the bottom of the potential well. The second case is a resonant bifurcation of periodic orbits with period 3 of the basic family of the system. Finally, the third case is a subcritical pitchfork bifurcation of the basic family that is connected with two saddle-node bifurcations. The stability diagram gives the variation of the H\'enon stability parameter $\alpha$ (see the appendix of \cite{katsanikas2018phase}) of a family of periodic orbits versus a parameter of the system (in our case, the energy, see \cite{contopoulos2004order}). We refer to the curve in the stability diagram representing the evolution of the H\'enon stability parameter of a family of periodic orbits versus the energy as the stability curve (see appendix \ref{appendix} for more details). 

In this section, we describe the three cases of bifurcations of periodic orbits using stability diagrams and LDs:

\subsection{Case I}  
\label{sub.1}

\begin{figure}[htbp]
    \centering
    \includegraphics[scale=0.35]{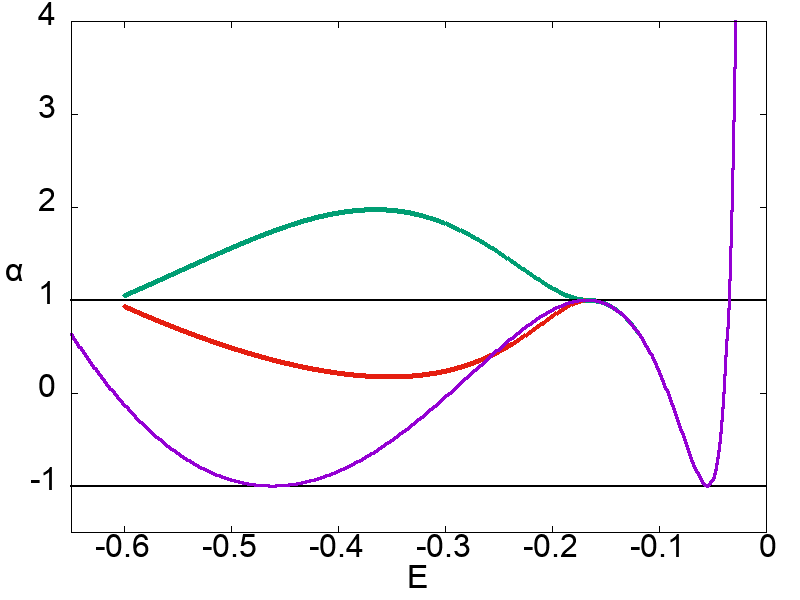}
    \caption{The stability diagram (the H\'enon stability parameter versus energy)  of periodic orbits. The red and green curves represent the stability curve for two stable families of periodic orbits with period 2 (A1 and A2) and for two unstable families (B1 and B2) of periodic orbits with period 2, respectively. The violet curve represents the stability curve for the family of the well, which is described two times,  and it generates the resonant families (with red and green colour respectively) when the curve is tangent with the axis $\alpha=1$.}
    \label{stab1}
\end{figure}

\begin{figure}[htbp]
    \centering
    A)\includegraphics[scale=0.4]{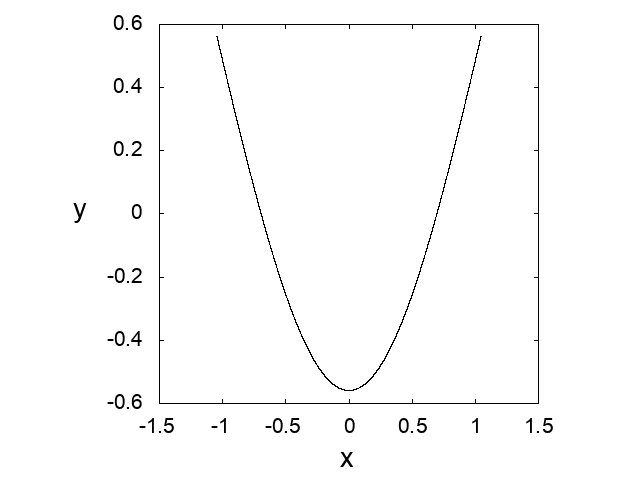}
    B)\includegraphics[scale=0.4]{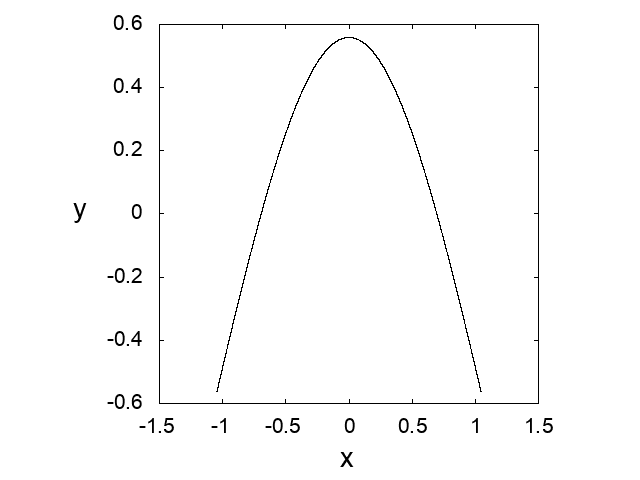}\\
    C)\includegraphics[scale=0.4]{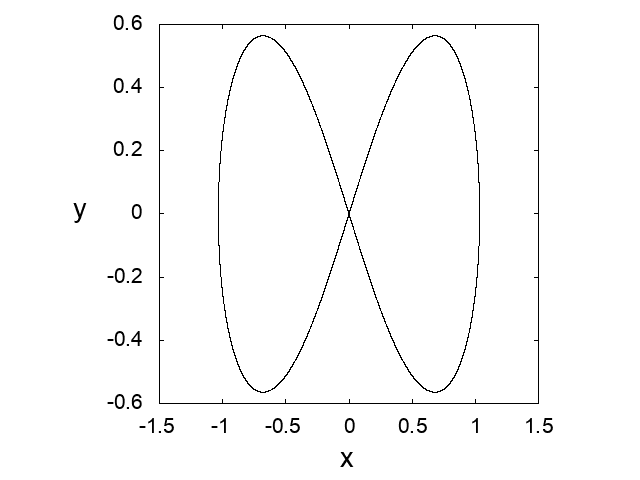}
    D)\includegraphics[scale=0.4]{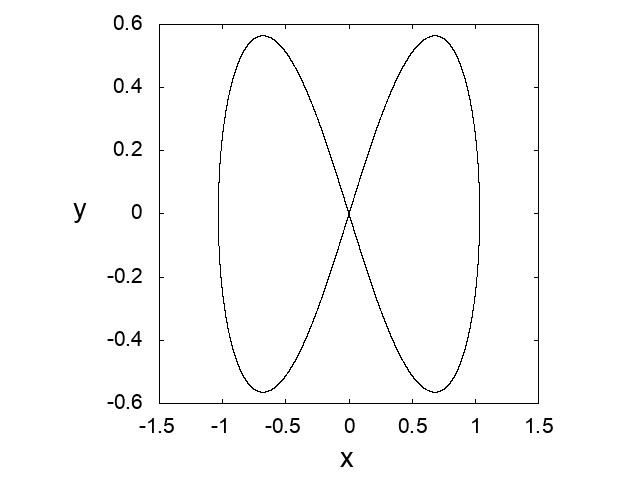}
    \caption{Periodic orbits with period 2 (for $E=-0.55$) in the configuration space. The periodic orbits in panels A) and B) are members of the two stable families (A1 and A2) with period 2 and the periodic orbits in panels C) and D) are members of the two unstable families (B1 and B2) with period 2. }
    \label{po-2}
\end{figure}

\begin{figure}[htbp]
    \centering
    \includegraphics[scale=0.45]{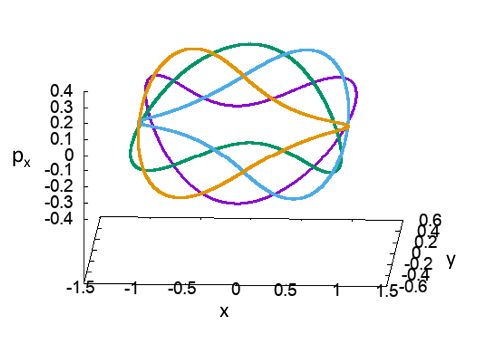}
    
    \caption{ The 3D projection of periodic orbits with period 2  (for $E=-0.55$) in the $(x,y,p_x)$ space. The periodic orbits are members  of the two stable families (A1 and A2 with violet and green colour respectively) with period 2 and members of the two unstable families (B1 and B2 with cyan and orange colour respectively) with period 2.  }
    \label{po-3d1}
\end{figure}

In this subsection, we describe the stability diagram and LDs for the first case of bifurcations in our system. The bifurcation for case I occurs at $E=-0.17$. This happens when the stability curve of the family of the well that is described has a tangency with the axis $\alpha=1$, see Fig.\ref{stab1}. At this point,  we have a pair of new families with period 2 that bifurcate from the family of the well. For more details about this kind of bifurcation, see the appendix \ref{appendix}. In our case, we have actually two pairs of new families of periodic orbits with period 2 because of the symmetry of the potential.  Every pair has one stable family and one unstable family. One pair consists of the A1 and B1 families, and the other consists of the A2 and B2 families. These families are resonant bifurcations of the family of the well. This kind of bifurcation is of inverse type. This means that the new bifurcations that were born continue for lower values of energy than that of the bifurcation, see Fig.\ref{stab1}. The families A1 and A2 are stable, and B1 and B2 are unstable, see Fig.\ref{stab1}. Geometrically, the periodic orbits A1 and A2 projected onto configuration space are concave upward and downward curves respectively, see for example in Fig. \ref{po-2}. The periodic orbits of B1 and B2 families are represented by a horizontal eight figure in the configuration space, see Fig.\ref{po-2}. An example of the 3D representation of  the periodic orbits in the space $(x,y,p_x)$ is depicted in Fig. \ref{po-3d1}. 

\begin{figure}[htbp]
\centering
%    A)urcation p\includegraphics[width=0.7\textwidth]{LD_action_total_x-px_tau_500_k_kc_E_-0_65.png}
%    \textcolor{red}{B)}\includegraphics[width=0.7\textwidth]{LD_action_total_x-px_tau_500_k_kc_E_-0_61.png}
    % A)\includegraphics[width=0.85\textwidth]{lds-055-new.png}
    A)\includegraphics[page=1,trim=0cm 1.25cm 0cm 1.25cm, clip=true, width=0.8\textwidth]{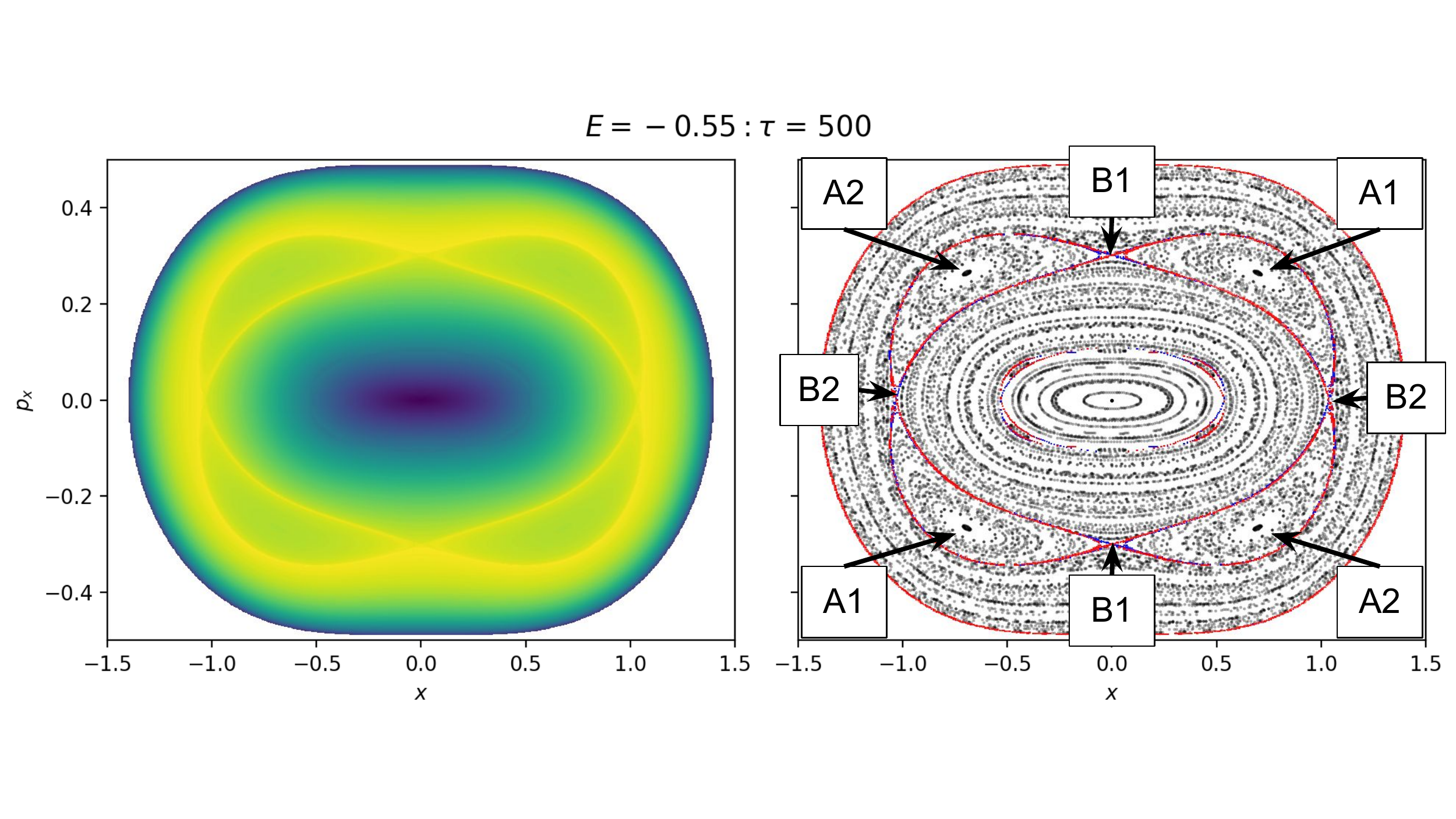}
    B)\includegraphics[page=2,trim=0cm 1.25cm 0cm 1.25cm, clip=true, width=0.8\textwidth]{LD-PP_figures.pdf}
    \caption{
    Different stages in the evolution of bifurcating POs as the energy of the system, $E$, increases away from the non-hyperbolic equilibrium point.
    Action-based Lagrangian Descriptor (Left) and Poincar\'e surfaces of section (Right) maps are shown for the phase space $x-p_x$ section for $E =-0.55, -0.462$ (Top to Bottom). Invariant stable (Blue) and unstable (Red) manifolds bounding POs are overlaid in every Poincar\'e surface of section map. We indicate the positions of the periodic orbits of families A1, A2, B1 and B2 in Poincar\'e section for $E=-0.55$. Similarly, we have the positions of the periodic orbits of these families for $E=-0.462$}.
    \label{fig:po-bifurcation_lds_1}
\end{figure}

Lagrangian descriptors can reveal the phase space structures and especially the invariant manifolds associated with the unstable periodic orbits. On the other hand, the classical method of Poincar\'e sections can reveal easily the structure of the KAM tori around the stable periodic orbits. We will see in this subsection how the combination of these two methods can detect the resonant bifurcations of the families of the family of the well (families of periodic orbits with period 2). In the panel A of Fig. \ref{fig:po-bifurcation_lds_1} is the Poincar\'e section for $E=-0.55$. We observe invariant curves around the stable periodic orbit of the family of the well that represents the KAM tori that exist around this periodic orbit, according to the KAM theorem \cite{kolmogorov1954,arnold1963,moser1962}. We also observe a chain of four islands that represents a resonance zone around the central periodic orbit of the family of the well. The centres of these islands give us the location of the stable periodic orbits of the families A1 and A2. Two of them correspond to the family A1 and the others to the family A2, see panel A of Fig. \ref{fig:po-bifurcation_lds_1}. These points can be used as an initial guess to a continuation method for periodic orbits and can give us the families A1 and A2 that we see in Fig.\ref{stab1}. Poincar\'e map is a classic method for finding an initial guess to compute stable families of periodic orbits. However, it is not always easy to find the unstable families of periodic orbits using the information of the Poincar\'e map. For this reason, we use the LDs in order to reveal the location of the unstable periodic orbits. In Fig.\ref{fig:po-bifurcation_lds_1}, we computed the invariant manifolds using LDs, and observe a typical picture of a resonance zone with four islands and the invariant manifolds that form lobes around these islands. This picture gives us the location of the unstable periodic orbits of B1 and B2 families that are between these lobes, see the points that are indicated by arrows in Fig. \ref{fig:po-bifurcation_lds_1}. This means that the LDs give us the location of the unstable periodic orbits of the families B1 and B2 (the unstable resonant bifurcating families with period 2). We can use these unstable periodic orbits as a good initial guess to a continuation method of periodic orbits and can give us the families B1 and B2 that we see in Fig.\ref{stab1}. We can continue this procedure and find the location of the periodic orbits for this resonance zone (with the four islands and the associated unstable periodic orbits) for larger values of energy (for example, $E=-0.462$, see the resonance zone with the four islands around the stable periodic orbits of A1 and A2 and the invariant manifolds of the associated unstable periodic orbits of B1 and B2 in the panel B of  Fig.\ref{fig:po-bifurcation_lds_1}). This becomes more difficult as we increase the value of the energy and we approach the bifurcation point $(E=-0.17)$ because the islands and the associated lobes that are formed from the invariant manifolds of the associated unstable periodic orbits become very small. The largest value of energy for which we can find the location of the unstable periodic orbits of B1 and B2 using LDs is $E=-0.2$ (that is lower by $0.03$ from the energy of the bifurcation point). This means that we start to detect the stable and the unstable periodic orbits of the resonant bifurcating families for values of energy very close to the bifurcation.    

\subsection{Case II}
\label{sub.2}

\begin{figure}[htbp]
    \centering
    \includegraphics[scale=0.4]{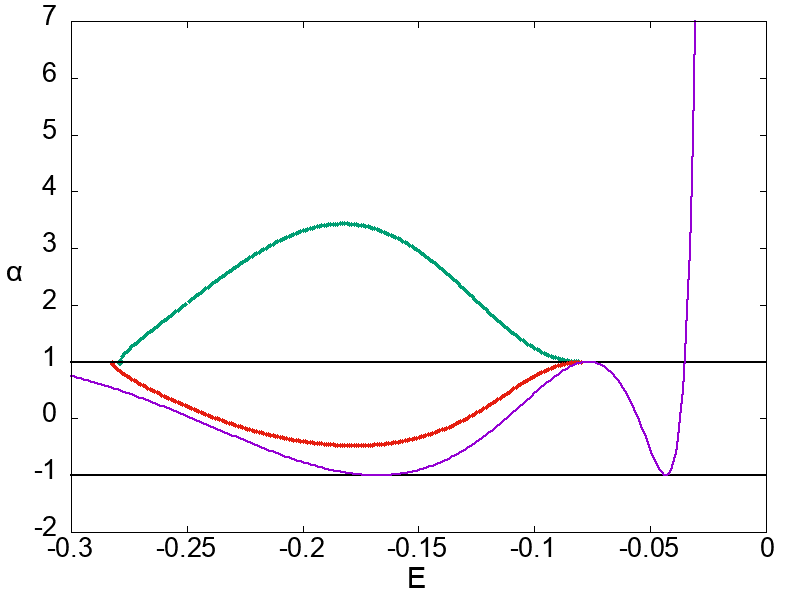}
    \caption{The stability diagram (the H\'enon stability parameter versus energy)  of periodic orbits. The red and green curves represent the stability curve for two stable families (A3 and A4) of periodic orbits with period 3  and for two unstable families (B3 and B4) of periodic orbits with period 3, respectively. The violet curve represents the stability curve for the basic family of the well, which is described three times,  and it generates the resonant families (with red and green colour) when the curve is tangent with the axis $\alpha=1$.}
    \label{stab2}
\end{figure}

\begin{figure}[htbp]
    \centering
    A)\includegraphics[scale=0.42]{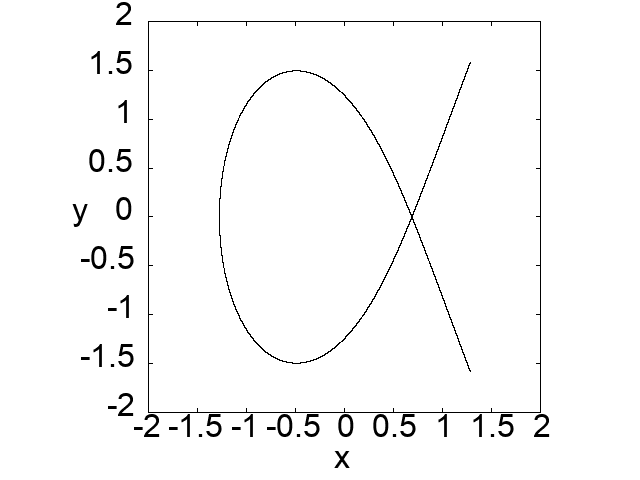}
    B)\includegraphics[scale=0.42]{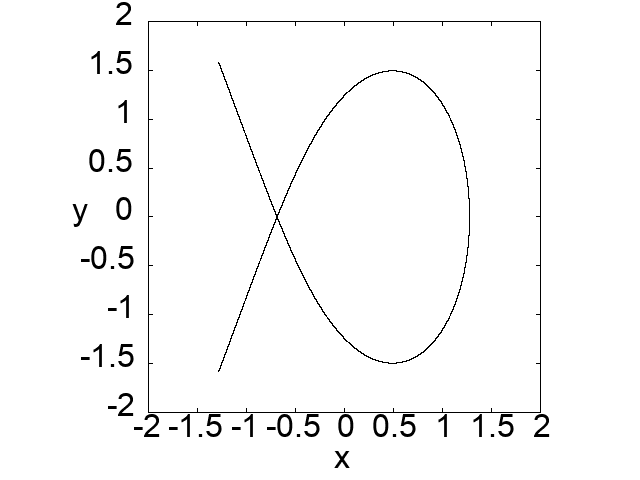}\\
    C)\includegraphics[scale=0.42]{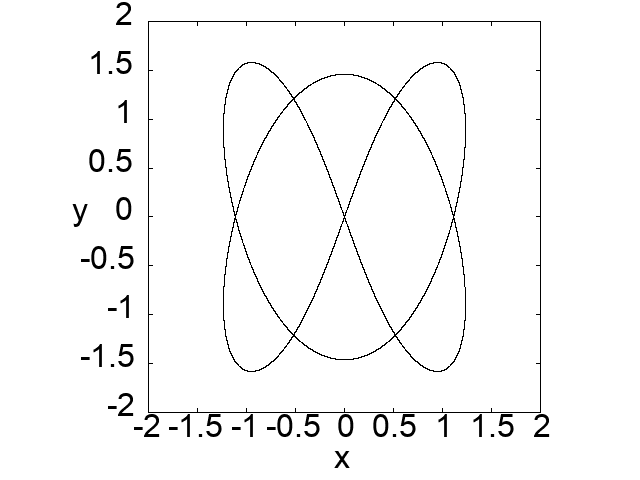}
    D)\includegraphics[scale=0.42]{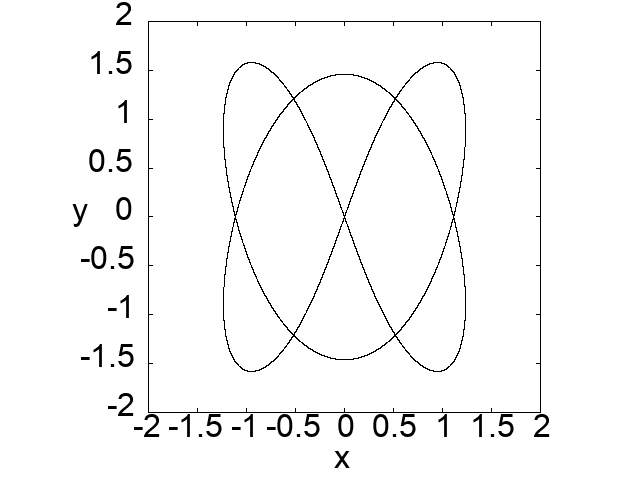}
    \caption{Periodic orbits with period 3  (for $E=-0.25$) in the configuration space. The periodic orbits in panels A) and B)  are members  of the two stable families (A3 and A4) with period 3 and the periodic orbits in panels C) and D) are members of the two unstable families (B3 and B4) with period 3. }
    \label{po-3}
\end{figure}

\begin{figure}[htbp]
    \centering
    A)\includegraphics[scale=0.45]{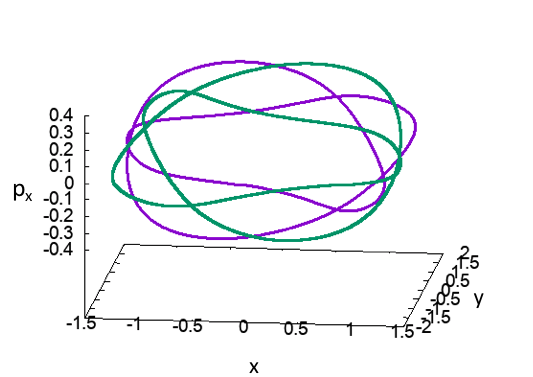}
    B)\includegraphics[scale=0.5]{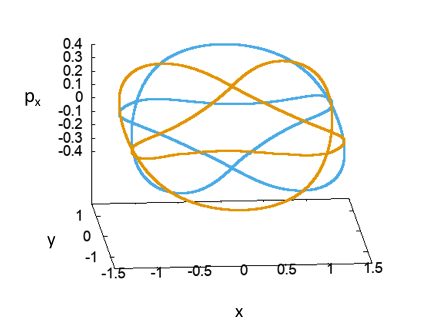}
    \caption{ The 3D projection of periodic orbits with period 3 (for $E=-0.25$) in the $(x,y,p_x)$ space. The periodic orbits are members  of the two stable families (A3 and A4 with violet and green color respectively in the panel A) with period 3 and members of the two unstable families (B3 and B4 with cyan and orange color respectively in the panel B) with period 3.  }
    \label{po-3d2}
\end{figure}

In this subsection, we describe the stability diagram and LDs for the second case of bifurcations in our system. The bifurcation for the case II occurs at $E=-0.075$. As in case I, we have one resonant bifurcation of the family of the well. The only difference with the case I is that now the two pairs of the bifurcating families are families of periodic orbits with period 3. This means that we have two pairs (each pair has one stable family and one unstable) of bifurcating families (the one pair is A3 and B3 and the other A4 and B4). The families A3 and A4 are stable, and the families B3 and B4 are unstable, see Fig. \ref{stab2}. This bifurcation occurs because of the tangency of the stability curve of the family of the well that was described three times on the Poincar\'e  section ($y=0$ with $p_y>0$) with the axis $\alpha=1$, see Fig. \ref{stab2},  at $E=-0.075$. The periodic orbits of A3 and A4 are represented as a right fish type curve and a left fish type curve, respectively, in the configurations space, see Fig. \ref{po-3}. The periodic orbits of B3 and B4 families are represented by a closed curve with many intersections in the configuration space, see  Fig. \ref{po-3}. An example of the  3D representation of  the periodic orbits  in the space $(x,y,p_x)$ is depicted in Fig. \ref{po-3d2}. 

\begin{figure}[htbp]
    \centering
    A)\includegraphics[page=3,trim=0cm 1.25cm 0cm 1.25cm, clip=true,width=0.8\textwidth]{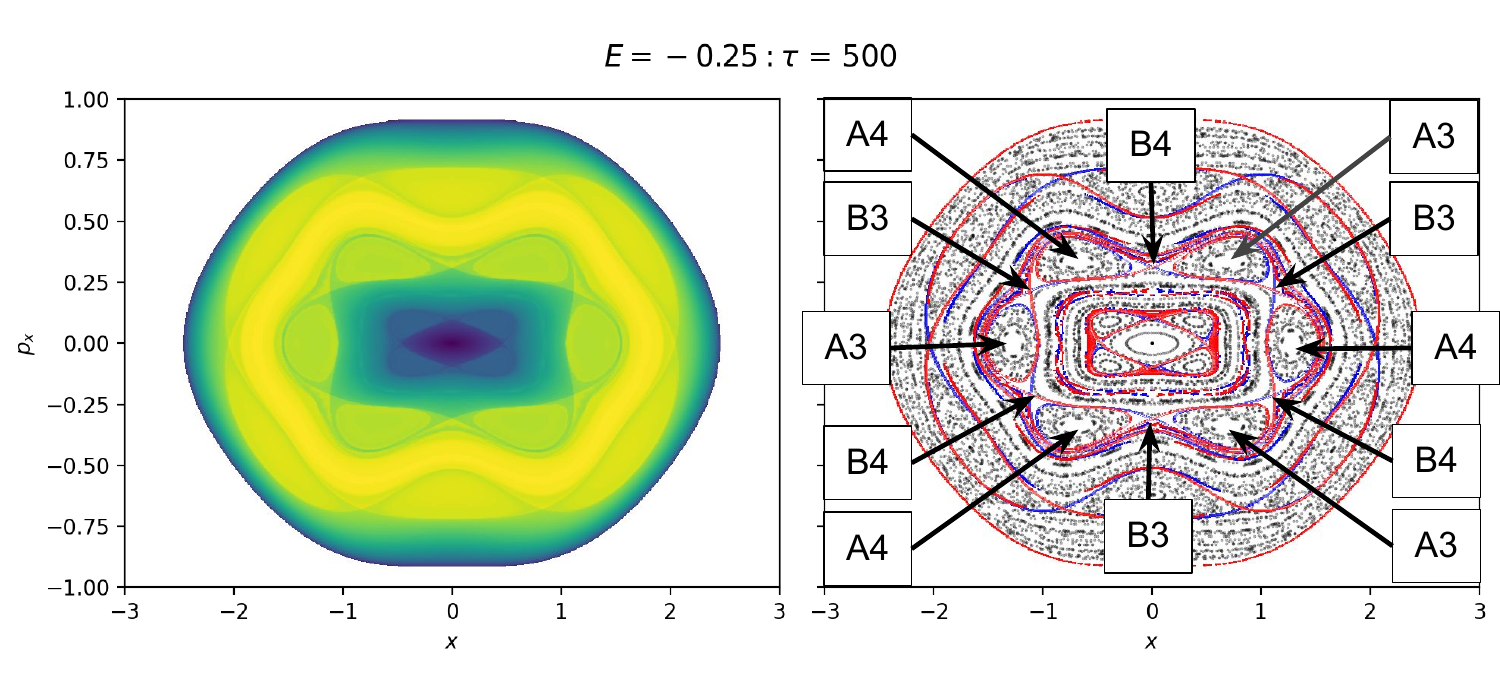}
    B)\includegraphics[page=4,trim=0cm 1.25cm 0cm 1.25cm, clip=true,width=0.8\textwidth]{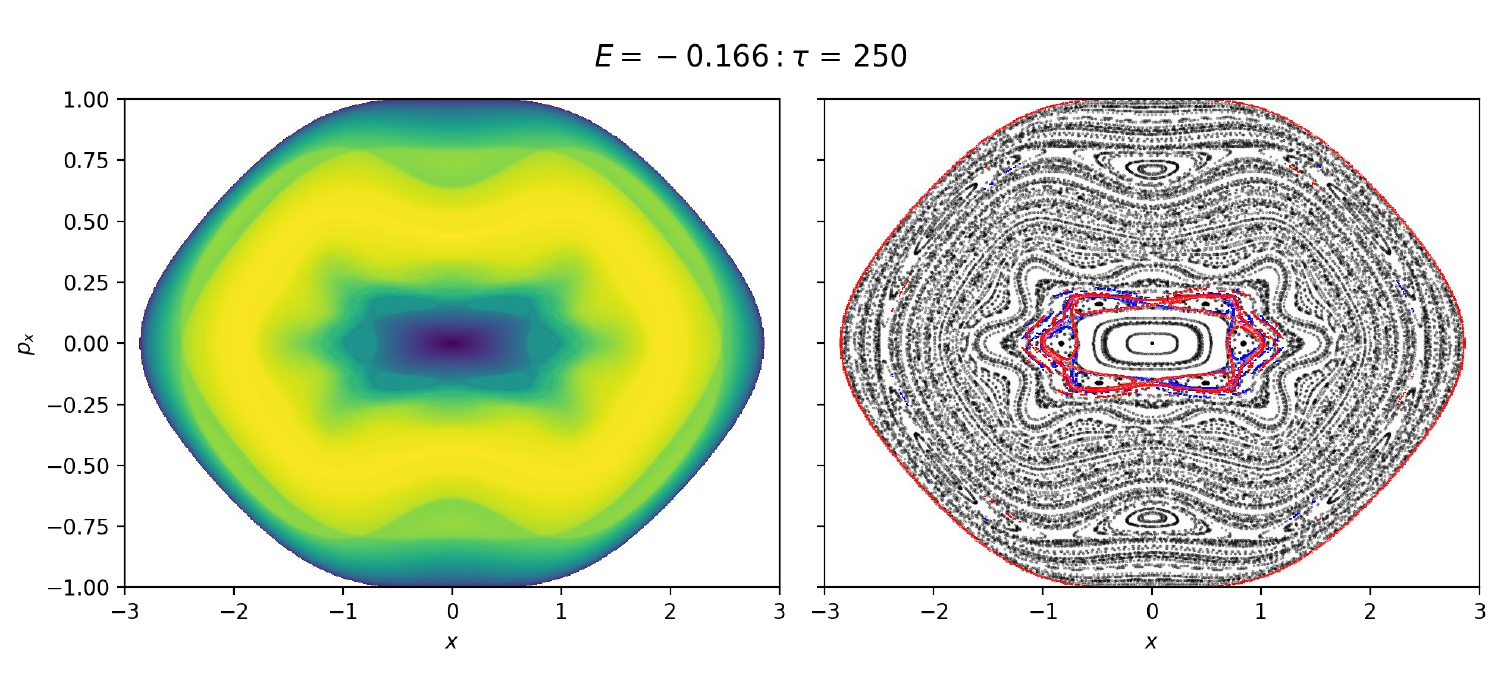}
    % C)\includegraphics[width=0.7\textwidth]{LD_total_x-px_tau_250_k_kc_E_-0_15.png}
    % D)\includegraphics[width=0.7\textwidth]{LD_total_x-px_tau_250_k_kc_E_-0_1.png}
    \caption{
    Same as in the previous figure but for energies $-0.25$ and $-0.166$ (\emph{Top to Bottom}). We indicate the positions of the periodic orbits of families A3, A4, B3 and B4 in Poincar\'e section for $E=-0.25$. Similarly, we have the positions of the periodic orbits of these families for $E=-0.166$.}
    \label{fig:po-bifurcation_lds_2}
\end{figure}

In this case, as we saw in case I (see subsection \ref{sub.1}), the combination of the method of LDs with the method of Poincar\'e sections can detect the resonant bifurcations of the families of the family of the well. The only difference is that the bifurcating families are families of periodic orbits with period 3 (instead of 2, as in the case I). In the panel A of Fig. \ref{fig:po-bifurcation_lds_2} we see the Poincar\'e section  for $E=-0.25$. We observe invariant curves around the stable periodic orbit of the family of the well and a chain of six islands that represents a resonance zone around the central periodic orbit of the family of the well. The centres of these islands give us the location of the stable periodic orbits of the families A3 and A4. The three of them correspond to the family A3 and the others to the family A4 (see panel A of Fig. \ref{fig:po-bifurcation_lds_2}). In Fig. \ref{fig:po-bifurcation_lds_2}, we computed the invariant manifolds using LDs, and we observe a typical picture of a resonance zone with six islands and the invariant manifolds that form lobes around these islands. This picture gives us the location of the unstable periodic orbits of B3 and B4 families that are between these lobes (see the points that are indicated by arrows in Fig. \ref{fig:po-bifurcation_lds_2}). As we saw in case I, Poincar\'e sections and LDs can give us the location of the stable and unstable periodic orbits, respectively. These periodic orbits can be used as an initial guess to a well-established method of the continuation of periodic orbits and can give us the bifurcating families A3, A4, B3 and B4 that we see in Fig.\ref{stab2}. As we can see in Fig.\ref{fig:po-bifurcation_lds_2}, we can find the location of the periodic orbits for this resonance zone (with the six islands and the associated unstable periodic orbits) for larger values of energy (for example, $E=-0.166$, see the resonance zone with the six islands around the stable periodic orbits of A3 and A4 and the invariant manifolds of the associated unstable periodic orbits of B3 and B4 in the panel B of  Fig.\ref{fig:po-bifurcation_lds_2}). This becomes more difficult as we increase the value of the energy and we approach the bifurcation point $(E=-0.075)$ because the islands and the associated lobes that are formed from the invariant manifolds of the associated unstable periodic orbits become very small. The largest value of energy for which we can find the location of the unstable periodic orbits of B3 and B4 using LDs is $E=-0.166$ (that is lower by $0.091$ from the energy of the bifurcation point). 

\subsection{Case III}
\label {sub.3}

\begin{figure}[htbp]
    \centering
    \includegraphics[scale=0.4]{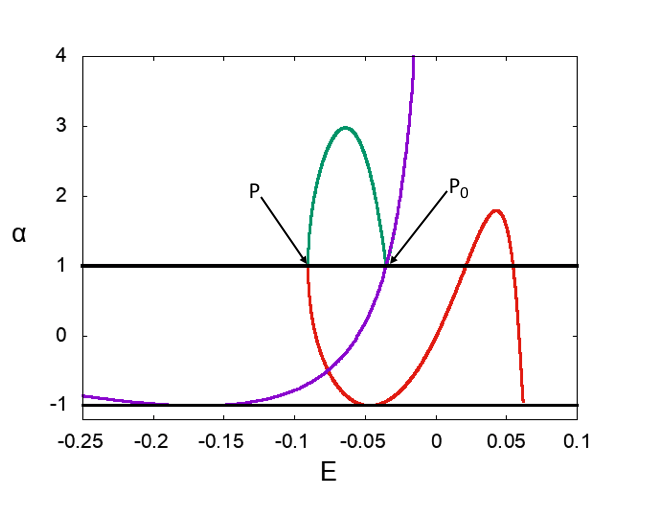}
    \caption{The stability diagram (the H\'enon stability parameter versus energy) of periodic orbits. The red and green curves represent the stability curve for two stable families of periodic orbits (A5 and A6) and for two unstable families (B5 and B6) of periodic orbits, respectively.  The violet curve represents the stability curve for the family of the well. We indicate by an arrow the point $P_0$ ($E=-0.036$) that corresponds to the point that we have a transition of the basic family of the well, from stability to instability. At this point, the periodic orbit has a subcritical pitchfork bifurcation and the generation of the unstable families B5 and B6 from the family of the well. Each of the families A5 and A6 is introduced in the system through a saddle-node bifurcation of the families B5 and B6, respectively at the point P (an arrow indicates that for $E=-0.0905$). }
    \label{stab-s}
\end{figure}

\begin{figure}[htbp]
    \centering
    \includegraphics[scale=1.6]{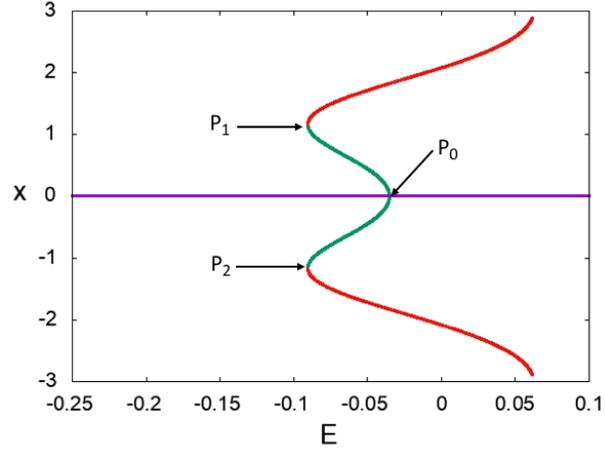}
    \caption{The characteristic (bifurcation) diagram (the position $x$ of the periodic orbits on the Poincar\'e section $y=0$ with $p_y>0$ versus energy) for the case of two saddle-node bifurcations and one subcritical pitchfork bifurcation of periodic orbits. The red and green curves represent the characteristic curves for two stable families of periodic orbits  (A5 for positive values $x$ and A6 for negative values $x$) and for two unstable families (B5 for positive values $x$ and B6 for negative values $x$) of periodic orbits respectively.  The violet curve represents the characteristic curve for the family of the well. We indicate by an arrow the point $P_0$ ($E=-0.036$) that corresponds to the point that we have a transition of the basic family of the well, from stability to instability (corresponds to the point $P_0$ in Fig. \ref{stab-s}). At this point, the periodic orbit undergoes a subcritical pitchfork bifurcation and the generation of the unstable families B5 and B6 from the family of the well. Each of the families A5 and A6 is introduced in the system through a saddle-node bifurcation of the families  B5 and B6 respectively at the points $P_1$ and $P_2$ (that are indicated by  arrows for $E=-0.0905$ - these points correspond to the point P in Fig.\ref{stab-s}).} 
    \label{char-s}
\end{figure}

\begin{figure}[htbp]
    \centering
    A)\includegraphics[scale=0.42]{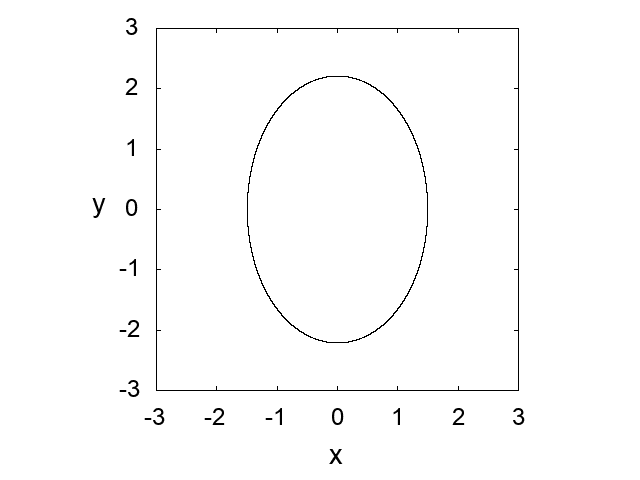}
    B)\includegraphics[scale=0.42]{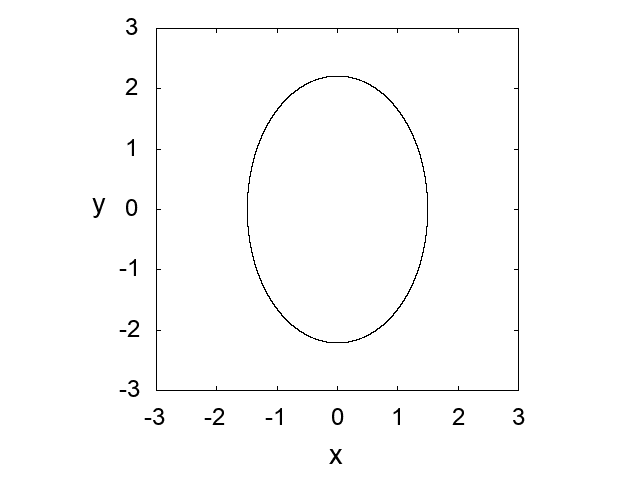}\\
    C)\includegraphics[scale=0.42]{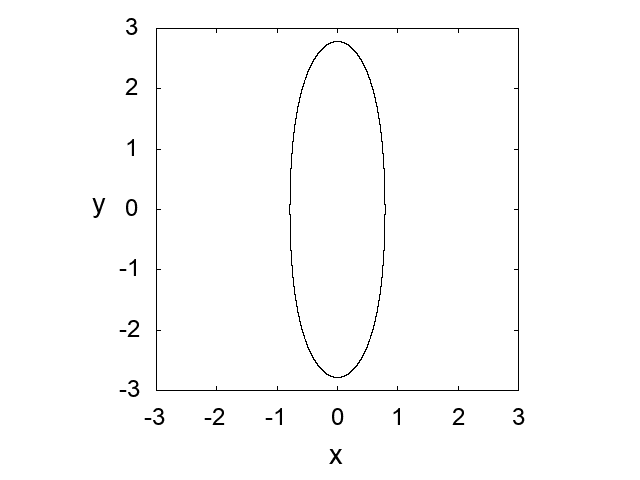}
    D)\includegraphics[scale=0.42]{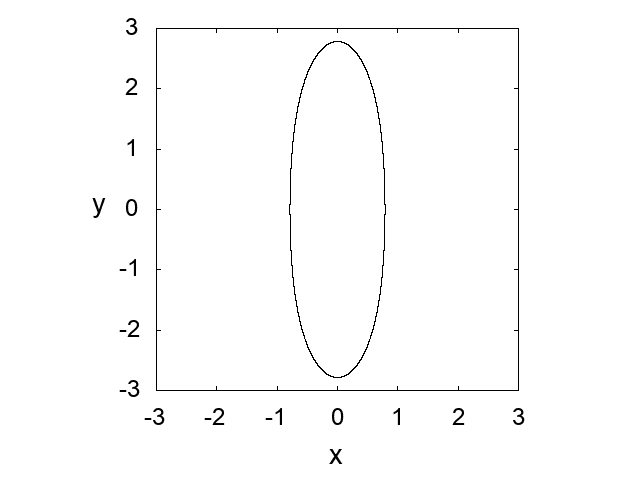}
    \caption{Periodic orbits  (for $E=-0.075$) in the configuration space. The periodic orbits in panels A) and B) are members of the two stable families (A5 and A6) and the periodic orbits in panels C) and D) are members of the two unstable families (B5 and B6).}
    \label{po-1}
\end{figure}

\begin{figure}[htbp]
    \centering
    \includegraphics[scale=0.45]{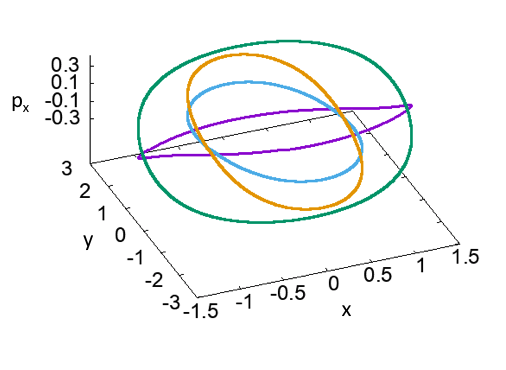}
    \caption{The 3D projection of periodic orbits (for $E=-0.075$) in the $(x,y,p_x)$ space. The periodic orbits are members of the two stable families (A5 and A6 with violet and green color respectively) and members of the two unstable families (B5 and B6 with cyan and orange color respectively).}
    \label{po-3d3}
\end{figure}

In this subsection, we investigate the third case of bifurcation using LDs and stability diagrams. In Fig. \ref{stab-s}, the stability curve of the stable families A5 and A6 (red curve) and of the unstable families B5 and B6 (with green colour) meet the axis $\alpha=1$ at the point P (for $E=-0.0905)$. The point P in the stability diagram corresponds to the symmetric points P1 and P2 of the characteristic diagram, see Fig. \ref{char-s}. The point P1 represents the minimum and maximum of the characteristic curves of the families A5 and B5, respectively. This means we have a saddle-node bifurcation for families A5 and B5 (see, for example the appendix of \cite{katsanikas2018phase} and \cite{contopoulos2004order}). The A5 family and B5 families represent the stable branch and unstable branch of this bifurcation, respectively. Similarly, the point P2 is the point of the saddle-node bifurcation for the families A6  and B6. The unstable branches of these bifurcations the families B5 and B6 continue for larger values of energy until they arrive at the point $P_0$ (for $E=-0.036$ - see the stability diagram in Fig.\ref{stab-s} and characteristic diagram in Fig.\ref{char-s}). At this point, the family of the wells becomes unstable (see Fig.\ref{stab-s}), and we see also the families B5 and B6 have their maximum value of energy at this point and exist on the stable part of the family of the wells, see Fig.\ref{char-s}. Consequently, the point $P_0$ is the point where we have a subcritical pitchfork bifurcation and families B5 and B6 are the bifurcating families of the family of the wells (see for more details about his kind of bifurcations in the appendix of \cite{katsanikas2018phase}). This means that the two saddle-node bifurcations at points P1 and P2 are connected with the subcritical pitchfork bifurcation at the point $P_0$, see Fig.\ref{char-s}. The periodic orbits of A5, A6, B5 and B6 are represented as ellipses in the configurations space, see for example  Fig.\ref{po-1}. The periodic orbits of the families A5 and A6 are more elongated in the $x$-axis than the periodic orbits of the families B5 and B6, see for example in Fig.\ref{po-1}. On the contrary, the periodic orbits of the families B5 and B6 are more elongated in the $y$-axis than the periodic orbits of A5 and A6 families, see for example in Fig. \ref{po-1}. An example of the 3D representation of  the periodic orbits in the space $(x,y,p_x)$ is depicted in Fig.\ref{po-3d3}.  

\begin{figure}[htbp]
    \centering
    A)\includegraphics[page=5,trim=0cm 1.25cm 0cm 1.25cm, clip=true,width=0.8\textwidth]{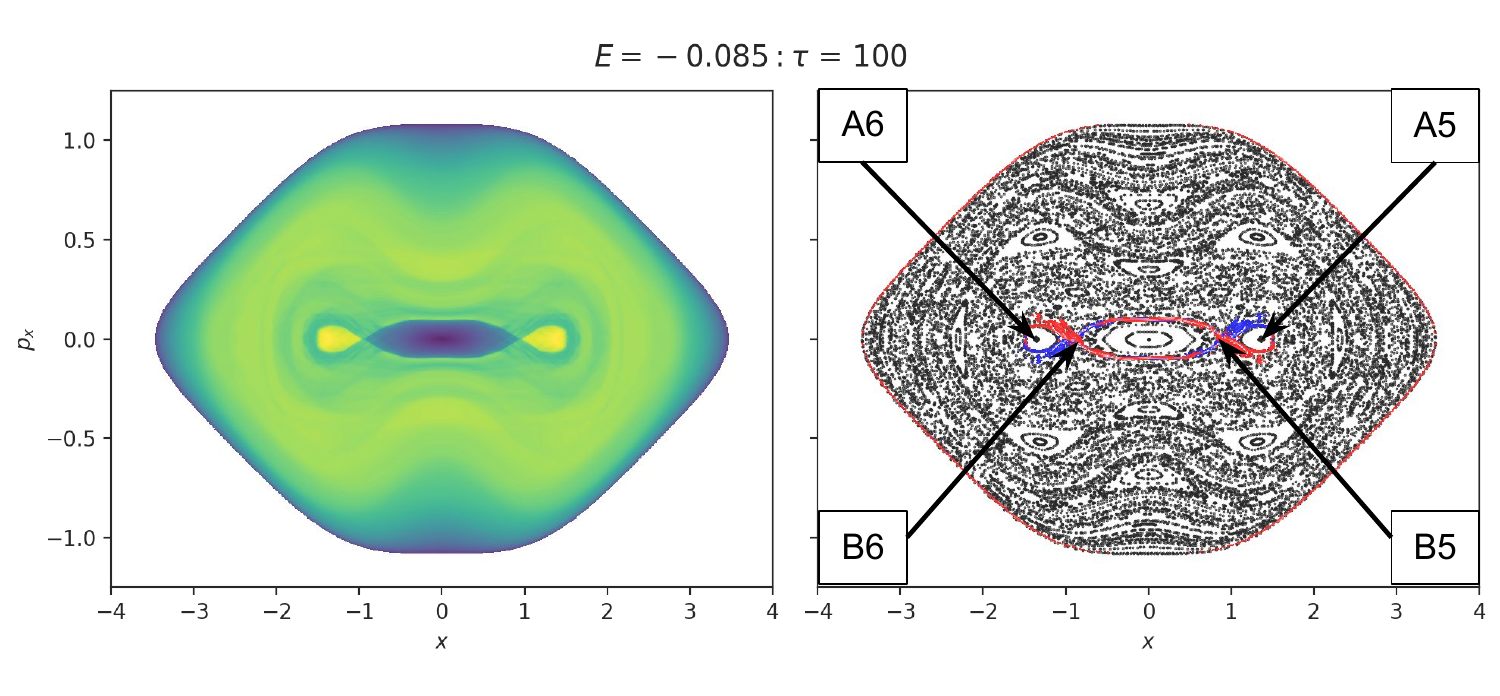}
    B)\includegraphics[page=7,trim=0cm 1.25cm 0cm 1.25cm, clip=true, width=0.8\textwidth]{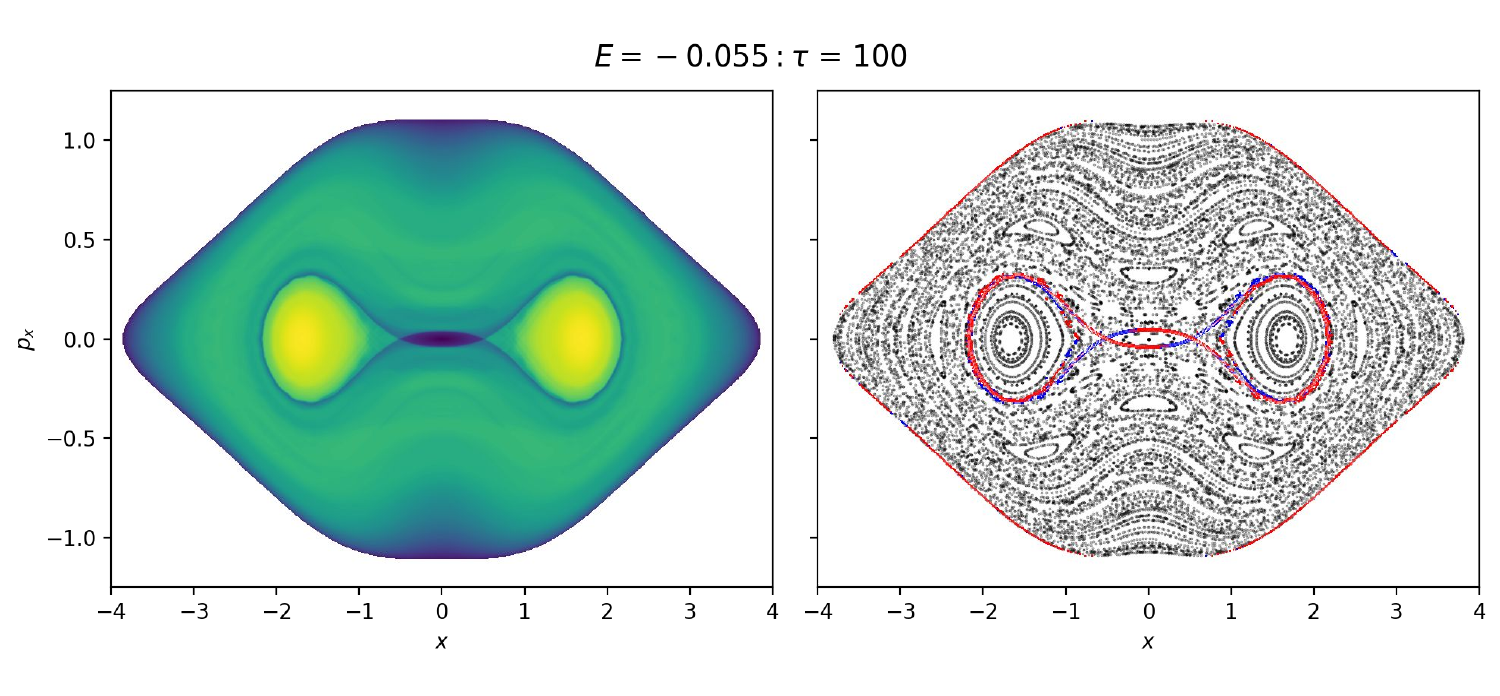}
    C)\includegraphics[page=8,trim=0cm 1.25cm 0cm 1.25cm, clip=true, width=0.8\textwidth]{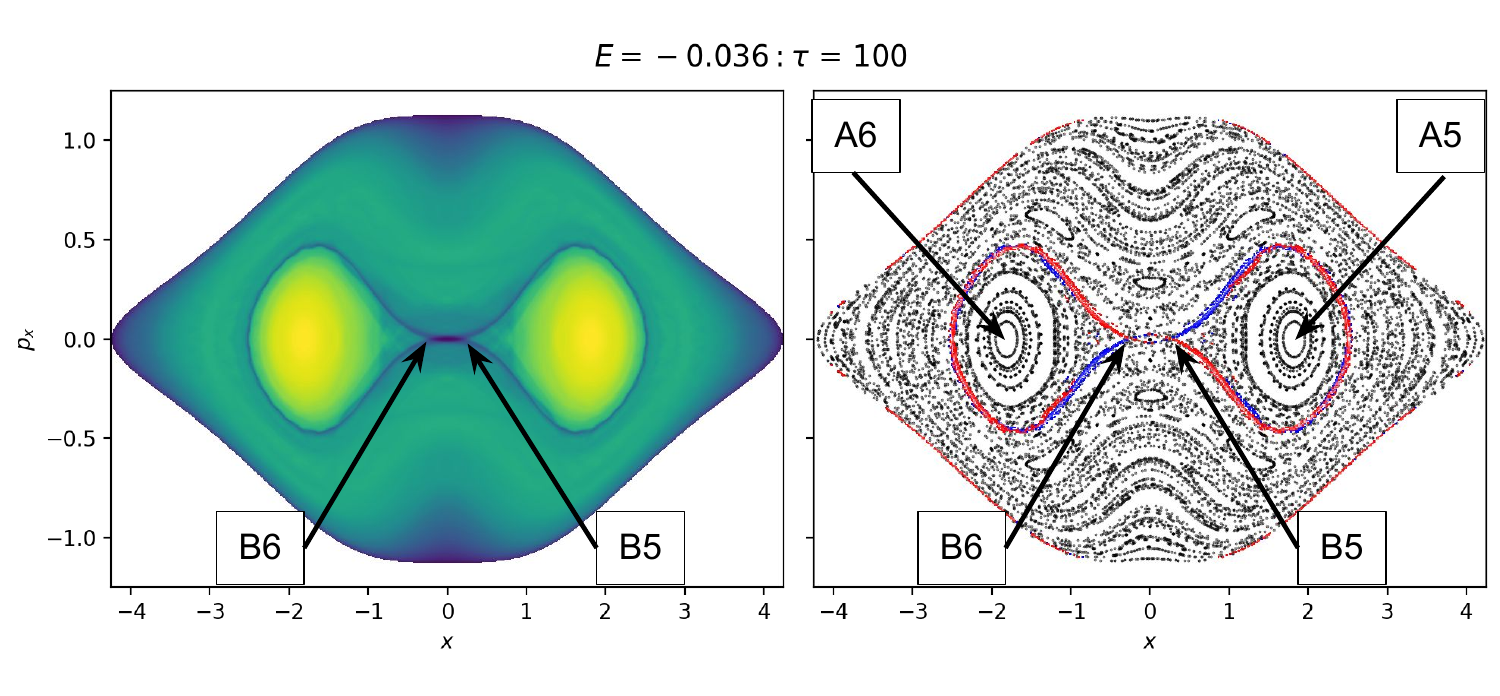}
    \caption{Same as in the previous figure but for energies $-0.075 \leq E \leq -0.036$ (\emph{Top to Bottom}). We indicate the positions of the periodic orbits of families A5, A6, B5 and B6 in Poincar\'e section for $E=-0.075$ and $E=-0.036$. Similarly, we have the positions of the periodic orbits of these families for $E=-0.055$.}
    \label{fig:po-bifurcation_lds_3}
\end{figure}

In Fig. \ref{fig:po-bifurcation_lds_3} we see the Poincar\'e sections can detect the periodic orbits of the stable families A5 and A6, that are the centres of the islands that are on both sides of the island that is around the stable periodic orbit of the family of the well at the centre, see the right column of the panels in Fig. \ref{fig:po-bifurcation_lds_3}.  As in the previous cases, the LDs detect the positions of the periodic orbits of the unstable families B5 and B6, see Fig.\ref{fig:po-bifurcation_lds_3}. Consequently, The combination of these two methods gives us the location of the periodic orbits of the bifurcating families of the saddle-node bifurcations (at the points P1 and P2 in Fig. \ref{char-s}) and of the subcritical pitchfork bifurcation (at the point $P_0$ in Fig. \ref{stab-s}) for a large interval of energy , see for example  Fig. \ref{fig:po-bifurcation_lds_3}. Poincar\'e sections give us the location of the periodic orbits of the stable branches of the saddle-node bifurcations, the families A5 and A6, even for values of energy ($E=-0.085$ - Fig. \ref{fig:po-bifurcation_lds_3} ) close to the energy of the bifurcating points P1 and P2 $(E=-0.0905)$. LDs detect for the same values of energy the periodic orbits of the unstable branches of the saddle-node bifurcations, the families B5 and B6 , see Fig. \ref{fig:po-bifurcation_lds_3}. As we described before, the families B5 and B6 are also the bifurcating families of the family of the well through a subcritical pitchfork bifurcation (at the point $P_0$ of Fig. \ref{char-s}). The remarkable thing is that LDs can detect the location of the periodic orbits of these families even at the value of energy $E=-0.036$ (see Fig. \ref{fig:po-bifurcation_lds_3}) for which we have the subcritical pitchfork bifurcation. This means that the LDs can reveal the location of the unstable periodic orbits of the bifurcating families, at the case of a subcritical pitchfork bifurcation, from the beginning (for example, in this case, from the bifurcating point $P_0$ in Fig. \ref{char-s}).

% \begin{figure*}
%     \centering
%     A)\includegraphics[width=0.75\textwidth]{LD_total_x-px_tau_500_k_kc_E_0_01.png}
%     B)\includegraphics[width=0.75\textwidth]{LD_action_total_x-px_tau_500_k_kc_E_0_02.png}
%     C)\includegraphics[width=0.75\textwidth]{LD_total_x-px_tau_500_k_kc_E_0_05.png}
%     D)\includegraphics[width=0.75\textwidth]{LD_total_x-px_tau_500_k_kc_E_0_2.png}
    
%     \caption{
%     Same as in the previous figure but for energies within the range $0.01 \leq E \leq 0.2$ (\emph{Top to Bottom}).
%     \emph{Stage 9}. (A-C) For energies above zero, we observe the persistence of the non-hyperbolic saddle that developed in Stage 8, while the lateral islands gradually shrink to vanish for energies above the dissociation energy eventually. Ultimately, the system transitions from chaotic to regular as POs are no longer captured since the system becomes more open as the energy rises further (D). 
%     }
%     \label{fig:po-bifurcation_lds_4}
% \end{figure*}

\section{Conclusions and remarks}
\label{sec:conclutions}

The method of Lagrangian descriptors is a useful tool to visualise the phase space structure and understand the dynamics. In particular, it is a simple, direct, technique based on the behaviour of the trajectories that allows us to find unstable hyperbolic periodic orbits. Lagrangian descriptors easily detect the intersections between the stable and unstable manifolds. For a Hamiltonian system with two degrees of freedom, this intersection provides an initial guess to calculate the unstable hyperbolic periodic orbits with good accuracy. Using this initial guess it is then possible to follow the periodic orbits with a continuation method until their bifurcations. 

A simple methodology to study the bifurcation scenario is to carefully choose sets of initial conditions and calculate the Lagrangian descriptors and Poincar\'e maps. The information from the Lagrangian descriptor plots and Poincar\'e maps help us to find the hyperbolic and elliptic periodic orbits with good precision. This information is used to find the exact location of the periodic orbits using the  Newton-Raphson method. Then we are able to  follow the periodic orbits using  a continuation method as a parameter of the system is varied in order to find the entire family of periodic orbits and their bifurcations. 

In this paper, we investigated this capability for analyzing the bifurcations of the basic family (the family of stable periodic orbits at the bottom of the potential well associated to the pair of complex eigenvalues), and we do so as the energy of the system increases from the bottom of the well until the dissociation energy. The bifurcation scenario for this system is very rich, and it has two prominent resonant bifurcations, one associated with families of periodic orbits with period 2 and the other with families of periodic orbits with period 3. Also, there are two saddle-node bifurcations that are connected with one subcritical pitchfork bifurcation.

The main conclusions of this work are the following:

\begin{enumerate}

\item The Lagrangian descriptors can detect the unstable bifurcating families of resonant bifurcations even for values of energy close to the energy of the bifurcation point. In this case, the difference between the value of bifurcation energy obtained with a continuation method and the value calculated using Lagrangian descriptors is  $0.03$ for the case of families with period 2 and $0.091$ for the case of families with period 3.

\item The Lagrangian descriptors are capable of detecting the unstable branch of saddle-node bifurcations even for values of energy that differ by $0.0055$ from the energy of the bifurcation point. 

\item The Lagrangian descriptors can detect the periodic orbits of the unstable bifurcating families, at the case of a subcritical pitchfork bifurcation, from the starting point of these families (the value of energy that corresponds to the bifurcation point). 

\end{enumerate}

\section*{Acknowledgments}
The authors  acknowledge the financial support provided by the EPSRC Grant No. EP/P021123/1 and the Office of Naval Research Grant No. N00014-01-1-0769.

\appendix

\section{Linear Stability Analysis of the Equilibrium Points of Hamilton's Equations at the Bifurcation of the PES}
\label{appx:A}

Hamilton's equations that determine the systems dynamical evolution are given by:

\begin{equation}
\begin{cases}
\dot{x} = \dfrac{\partial H}{\partial p_x} = \dfrac{p_x}{m_1} \\[.5cm]
\dot{y} = \dfrac{\partial H}{\partial p_y} = \dfrac{p_y}{m_2} \\
\dot{p}_x = - \dfrac{\partial H}{\partial x} = - \dfrac{\partial V}{\partial x} = - 6 W_0 \, k^6 \left[\dfrac{x - d}{\left(\left(x - d\right)^2 + y^2 + k^2\right)^4} + \dfrac{x + d}{\left(\left(x + d\right)^2 + y^2 + k^2\right)^4}\right] \\[.8cm]
\dot{p}_y = - \dfrac{\partial H}{\partial y} = - \dfrac{\partial V}{\partial y} = - 6 W_0 \, k^6 y \left[\dfrac{1}{\left(\left(x - d\right)^2 + y^2 + k^2\right)^4} + \dfrac{1}{\left(\left(x + d\right)^2 + y^2 + k^2\right)^4}\right]
\end{cases}
\label{eq:ham_eq}
\end{equation}

In this work we consider the situation where both DoF have unit mass, that is $m_1 = m_2 = 1$. The equilibrium points $\mathbf{x}_e = (x_e,y_e,p_{x,e},p_{y,e})$ of the dynamical system given above are contained in configuration space, since $p_{x,e} = p_{y,e} = 0$, and they are also critical points of the PES, that is $\nabla V (x_e,y_e) = \mathbf{0}$. It is straightforward to check that under these conditions, the points have to satisfy:

\begin{equation}
y_e = 0 \quad , \quad \left(x_e - d\right) \left[\left(x_e + d\right)^2 + k^2\right]^4 + \left(x_e + d\right) \left[\left(x_e - d\right)^2 + k^2\right]^4 = 0 
\label{eq:eq_pts}
\end{equation}

\noindent
It is straightforward to show that $x_e = 0$ is always a solution to the equation above, and therefore the origin is always an equilibrium point of Hamilton's equations for this system. Moreover, there is also an equilibrium point located at infinity. Regarding the presence of other equilibria, this depends critically on the model parameters, and if they exist, their coordinates are of the form $\mathbf{x}_e = (x_e(k,d),0,0,0)$. 

The linear stability of these equilibrium points is determined by the eigenvalues of the Jacobian matrix given by the linearization of Eq. \eqref{eq:ham_eq} about each of the points. In order to construct the Jacobian, we need to compute the second order partial derivatives that characterize the Hessian matrix of the PES:

\begin{equation}
\begin{cases}
\dfrac{\partial^{\,	2} V}{\partial x^2} = 6 \, W_0 \, k^6 \left[\dfrac{-7\left(x - d\right)^2 + y^2 + k^2}{\left(\left(x - d\right)^2 + y^2 + k^2\right)^5} + \dfrac{-7\left(x + d\right)^2 + y^2 + k^2}{\left(\left(x + d\right)^2 + y^2 + k^2\right)^5} \right] \\[.8cm]

\dfrac{\partial^{\,	2} V}{\partial y^2} = 6 \, W_0 \, k^6 \left[\dfrac{\left(x - d\right)^2 - 7 y^2 + k^2}{\left(\left(x - d\right)^2 + y^2 + k^2\right)^5} + \dfrac{\left(x + d\right)^2 - 7y^2 + k^2}{\left(\left(x + d\right)^2 + y^2 + k^2\right)^5} \right] \\[.8cm]

\dfrac{\partial^{\,	2} V}{\partial x \partial y} = - 48 \, W_0 \, k^6 \, y \left[\dfrac{x - d}{\left(\left(x - d\right)^2 + y^2 + k^2\right)^5} + \dfrac{x + d}{\left(\left(x + d\right)^2 + y^2 + k^2\right)^5}\right]
\end{cases}
\label{eq:second_pder}
\end{equation}
We now examine the Hessian matrix of the PES evaluated at the origin:

\begin{equation}
\text{Hess}_{V}(0,0) = \begin{pmatrix}
\dfrac{\partial^{\,	2} V}{\partial x^2}(0,0) & \dfrac{\partial^{\,	2} V}{\partial x \partial y}(0,0) \\[.4cm]
\dfrac{\partial^{\,	2} V}{\partial y \partial x}(0,0) & \dfrac{\partial^{\, 2} V}{\partial y^2}(0,0)
\end{pmatrix} = \begin{pmatrix}
\dfrac{12 W_0 \, k^6 \left(k^2 - 7 d^2\right)}{\left(d^2 + k^2\right)^5} & 0 \\[.5cm]
0 & \dfrac{12 W_0 \, k^6}{\left(d^2 + k^2\right)^4}
\end{pmatrix}
\end{equation}

\noindent
The linear stability of the origin is characterized by the nature of the eigenvalues of the Jacobian matrix:

\begin{equation}
J(0,0) = \begin{pmatrix}
\mathbb{O}_2 & \hspace{.2cm} \mathbb{I}_2 \hspace{.1cm} \\[.3cm]
-\text{Hess}_{V}(0,0) & \hspace{.2cm} \mathbb{O}_2 \hspace{.1cm}
\end{pmatrix} = \begin{pmatrix}
0 & 0 & \hspace{.5cm} 1 & \hspace{.5cm} 0 \hspace{.2cm} \\[.4cm]
0 & 0 & \hspace{.5cm} 0 & \hspace{.5cm} 1 \hspace{.2cm} \\[.4cm]
\dfrac{12 W_0 \, k^6 \left(7 d^2 - k^2\right)}{\left(d^2 + k^2\right)^5} & 0 & \hspace{.5cm} 0 & \hspace{.5cm} 0 \hspace{.2cm} \\[.4cm]
0 & -\dfrac{12 W_0 \, k^6}{\left(d^2 + k^2\right)^4} & \hspace{.5cm} 0 & \hspace{.5cm} 0 \hspace{.2cm} 
\end{pmatrix}
\end{equation}

\noindent
The eigenvalues $\xi$ are given by:

\begin{equation}
\xi_{1,2} = \pm \, \omega \, i \quad,\quad \xi_{3,4} = \pm \, \omega \, \sqrt{\dfrac{7 d^2 - k^2}{ d^2 + k^2}} = \pm \, \omega \, \sqrt{\dfrac{7 -  \eta^2}{1 + \eta^2}}
 \end{equation}
 
\noindent
where $\eta = k / d$ is the ratio between the half-width of the single van der Waals potential and its displacement from the origin, and the angular frequency of oscillation in the linear approximation is:

\begin{equation}
\omega = \dfrac{2 \, \sqrt{3 W_0} \, k^3}{\left(d^2 + k^2\right)^2} =  \dfrac{2\sqrt{3 W_0} \, \eta^4}{k\left(1 + \eta^2\right)^2}
\end{equation}

\noindent
Therefore, the origin is an index-1 saddle equilibrium point when the condition $7 d^2 \geq k^2$ holds, which is equivalent to $0 < \eta < \sqrt{7}$. However, for $\eta > \sqrt{7}$ the index-1 saddle turns into a potential well (a center equilibrium point). This indicates that a pitchfork bifurcation takes place in the PES at the origin at the critical value $\eta = \sqrt{7}$. For this parameter value, the eigenvalues of the equilibrium point at the origin become:

\begin{equation}
   \xi_{1,2} = \pm \, \omega \, i = \pm \dfrac{49 \sqrt{3 W_0}}{32k} \, i \quad,\quad \xi_{3,4} = 0
\end{equation}

\noindent
Notice that for $\eta = \sqrt{7}$ the equilibrium point at the origin has a pair of complex eigenvalues, and a double zero eigenvalue. This means that it is a degenerated equilibrium point (parabolic nature). The complex eigenvalues will give rise to a family of stable periodic orbits due to the Lyapunov subcenter manifold theorem \cite{liapounoff1907probleme,moser1958generalization, kelley1967liapounov}, We illustrate the bifurcation phenomenon we have described above in Fig. \ref{fig:double_vdw_PES} C). This is done by examining how the shape of the PES changes along the $x$-axis for different values of the model parameter $\eta$.

\section*{appendix B - Stability Diagrams and Resonant Bifurcations}
\label{appendix}

In the appendix of \cite{katsanikas2018phase} we described the computation of the monodromy matrix $M$  of a 2D Poincar\'e map for the case of a periodic orbit. This matrix is $2\times2$ matrix that satisfies the symplectic identity (see \cite{katsanikas2018phase}):
\begin{equation}
M=
 \begin{bmatrix}
a & b\\
c & d
\end{bmatrix}
  \end{equation}
It can  be shown that the characteristic equation  of $M$ is  $\lambda^2-(a+d)\lambda+1=0$ and the eigenvalues $\lambda_1,\lambda_2$ are the roots of this  equation. We define the quantity $\alpha=(a+d)/2$  to be the H{\'e}non stability parameter (In a previous paper \cite{katsanikas2018phase} we give this definition as the absolute value of this in order to distinguish easier the two types of stability of periodic orbits, but here in this paper we use the initial definition, see e.q. \cite{contopoulos2004order}). There are two types of periodic orbits. The first type are the stable periodic orbits that have two complex conjugate eigenvalues on the unit circle. The second type are the unstable periodic orbits that have two eigenvalues on the real axis and outside from the unit circle. The first type occurs for $|\alpha|<1$  and the second for $|\alpha|>1$.  

The stability diagram gives the variation of the {\bf H\'enon stability parameter} $\alpha$ (see the appendix of \cite{katsanikas2018phase}) of a family of periodic orbits versus a parameter of the system, in our case, the energy (see \cite{contopoulos2004order}). We refer to the curve in the stability diagram that represents the evolution of the H\'enon stability parameter of a family of periodic orbits versus the energy as the stability curve. When a stability curve crosses the axis $\alpha=1$ or $\alpha=-1$, the orbit has a change of  stability. If it crosses the axis $\alpha=1$, the periodic orbit has a pitchfork bifurcation, and if it crosses the axis $\alpha=-1$, we have a period-doubling bifurcation \cite{contopoulos2004order}. In the case that the stability curve does not cross the axis $\alpha=1$  or $\alpha=-1$, but it becomes tangent with one of them, we have bifurcating families that appear in pairs (one stable and one unstable). If the tangency is with the axis $\alpha=-1$ or $\alpha=1$, the bifurcating families are of double or equal period, respectively. These bifurcations are referred to as  {\bf resonant bifurcations}. The main characteristic of these bifurcations is that the periodic orbits of the initial family do not change stability. If one family intersects the Poincar\'e section $n$ times and the stability curve has a tangency with the axis  $\alpha=1$, the bifurcating families have period $n$, see \cite{contopoulos2004order} page 105.

\bibliography{cirque}

\begin{thebibliography}{43}%
\makeatletter
\providecommand \@ifxundefined [1]{%
 \@ifx{#1\undefined}
}%
\providecommand \@ifnum [1]{%
 \ifnum #1\expandafter \@firstoftwo
 \else \expandafter \@secondoftwo
 \fi
}%
\providecommand \@ifx [1]{%
 \ifx #1\expandafter \@firstoftwo
 \else \expandafter \@secondoftwo
 \fi
}%
\providecommand \natexlab [1]{#1}%
\providecommand \enquote  [1]{``#1''}%
\providecommand \bibnamefont  [1]{#1}%
\providecommand \bibfnamefont [1]{#1}%
\providecommand \citenamefont [1]{#1}%
\providecommand \href@noop [0]{\@secondoftwo}%
\providecommand \href [0]{\begingroup \@sanitize@url \@href}%
\providecommand \@href[1]{\@@startlink{#1}\@@href}%
\providecommand \@@href[1]{\endgroup#1\@@endlink}%
\providecommand \@sanitize@url [0]{\catcode `\\12\catcode `\$12\catcode
  `\&12\catcode `\#12\catcode `\^12\catcode `\_12\catcode `\%12\relax}%
\providecommand \@@startlink[1]{}%
\providecommand \@@endlink[0]{}%
\providecommand \url  [0]{\begingroup\@sanitize@url \@url }%
\providecommand \@url [1]{\endgroup\@href {#1}{\urlprefix }}%
\providecommand \urlprefix  [0]{URL }%
\providecommand \Eprint [0]{\href }%
\providecommand \doibase [0]{https://doi.org/}%
\providecommand \selectlanguage [0]{\@gobble}%
\providecommand \bibinfo  [0]{\@secondoftwo}%
\providecommand \bibfield  [0]{\@secondoftwo}%
\providecommand \translation [1]{[#1]}%
\providecommand \BibitemOpen [0]{}%
\providecommand \bibitemStop [0]{}%
\providecommand \bibitemNoStop [0]{.\EOS\space}%
\providecommand \EOS [0]{\spacefactor3000\relax}%
\providecommand \BibitemShut  [1]{\csname bibitem#1\endcsname}%
\let\auto@bib@innerbib\@empty
%</preamble>
\bibitem [{\citenamefont {I{\~n}arrea}\ \emph {et~al.}(2011)\citenamefont
  {I{\~n}arrea}, \citenamefont {Palaci{\'a}n}, \citenamefont {Pascual},\ and\
  \citenamefont {Salas}}]{inarrea2011bifurcations}%
  \BibitemOpen
  \bibfield  {author} {\bibinfo {author} {\bibfnamefont {M.}~\bibnamefont
  {I{\~n}arrea}}, \bibinfo {author} {\bibfnamefont {J.~F.}\ \bibnamefont
  {Palaci{\'a}n}}, \bibinfo {author} {\bibfnamefont {A.~I.}\ \bibnamefont
  {Pascual}},\ and\ \bibinfo {author} {\bibfnamefont {J.~P.}\ \bibnamefont
  {Salas}},\ }\href {https://doi.org/10.1063/1.3600744} {\bibfield  {journal}
  {\bibinfo  {journal} {The Journal of chemical physics}\ }\textbf {\bibinfo
  {volume} {135}},\ \bibinfo {pages} {014110} (\bibinfo {year}
  {2011})}\BibitemShut {NoStop}%
\bibitem [{\citenamefont {Founargiotakis}\ \emph {et~al.}(1997)\citenamefont
  {Founargiotakis}, \citenamefont {Farantos}, \citenamefont {Skokos},\ and\
  \citenamefont {Contopoulos}}]{founargiotakis1997bifurcation}%
  \BibitemOpen
  \bibfield  {author} {\bibinfo {author} {\bibfnamefont {M.}~\bibnamefont
  {Founargiotakis}}, \bibinfo {author} {\bibfnamefont {S.}~\bibnamefont
  {Farantos}}, \bibinfo {author} {\bibfnamefont {C.}~\bibnamefont {Skokos}},\
  and\ \bibinfo {author} {\bibfnamefont {G.}~\bibnamefont {Contopoulos}},\
  }\href {https://doi.org/10.1016/S0009-2614(97)00931-7} {\bibfield  {journal}
  {\bibinfo  {journal} {Chemical physics letters}\ }\textbf {\bibinfo {volume}
  {277}},\ \bibinfo {pages} {456} (\bibinfo {year} {1997})}\BibitemShut
  {NoStop}%
\bibitem [{\citenamefont {Lega}\ \emph {et~al.}(2016)\citenamefont {Lega},
  \citenamefont {Guzzo},\ and\ \citenamefont {Froeschl{\'{e}}}}]{Lega2016}%
  \BibitemOpen
  \bibfield  {author} {\bibinfo {author} {\bibfnamefont {E.}~\bibnamefont
  {Lega}}, \bibinfo {author} {\bibfnamefont {M.}~\bibnamefont {Guzzo}},\ and\
  \bibinfo {author} {\bibfnamefont {C.}~\bibnamefont {Froeschl{\'{e}}}},\
  }\bibfield  {journal} {\bibinfo  {journal} {Lecture Notes in Physics}\ }\href
  {https://doi.org/10.1007/978-3-662-48410-4\_2} {10.1007/978-3-662-48410-4\_2}
  (\bibinfo {year} {2016})\BibitemShut {NoStop}%
\bibitem [{\citenamefont {Cincotta}\ and\ \citenamefont
  {Giordano}(2016)}]{Cicotta2016}%
  \BibitemOpen
  \bibfield  {author} {\bibinfo {author} {\bibfnamefont {P.~M.}\ \bibnamefont
  {Cincotta}}\ and\ \bibinfo {author} {\bibfnamefont {C.~M.}\ \bibnamefont
  {Giordano}},\ }\bibfield  {journal} {\bibinfo  {journal} {Lecture Notes in
  Physics}\ }\href {https://doi.org/10.1007/978-3-662-48410-4\_4}
  {10.1007/978-3-662-48410-4\_4} (\bibinfo {year} {2016})\BibitemShut {NoStop}%
\bibitem [{\citenamefont {Skokos}\ and\ \citenamefont {{Manos
  T.}}(2016)}]{Skokos2016}%
  \BibitemOpen
  \bibfield  {author} {\bibinfo {author} {\bibfnamefont {C.}~\bibnamefont
  {Skokos}}\ and\ \bibinfo {author} {\bibnamefont {{Manos T.}}},\ }\bibfield
  {journal} {\bibinfo  {journal} {Lecture Notes in Physics}\ }\href
  {https://doi.org/10.1007/978-3-662-48410-4\_5} {10.1007/978-3-662-48410-4\_5}
  (\bibinfo {year} {2016})\BibitemShut {NoStop}%
\bibitem [{\citenamefont {Gonzalez}\ and\ \citenamefont
  {Jung}(2012)}]{Gonzalez2012}%
  \BibitemOpen
  \bibfield  {author} {\bibinfo {author} {\bibfnamefont {F.}~\bibnamefont
  {Gonzalez}}\ and\ \bibinfo {author} {\bibfnamefont {C.}~\bibnamefont
  {Jung}},\ }\href {http://stacks.iop.org/1751-8121/45/i=26/a=265102}
  {\bibfield  {journal} {\bibinfo  {journal} {Journal of Physics A:
  Mathematical and Theoretical}\ }\textbf {\bibinfo {volume} {45}},\ \bibinfo
  {pages} {265102} (\bibinfo {year} {2012})}\BibitemShut {NoStop}%
\bibitem [{\citenamefont {Jim\'enez~Madrid}\ and\ \citenamefont
  {Mancho}(2009)}]{madrid}%
  \BibitemOpen
  \bibfield  {author} {\bibinfo {author} {\bibfnamefont {J.~A.}\ \bibnamefont
  {Jim\'enez~Madrid}}\ and\ \bibinfo {author} {\bibfnamefont {A.~M.}\
  \bibnamefont {Mancho}},\ }\href {https://doi.org/10.1063/1.3056050}
  {\bibfield  {journal} {\bibinfo  {journal} {Chaos}\ }\textbf {\bibinfo
  {volume} {19}},\ \bibinfo {pages} {013111} (\bibinfo {year}
  {2009})}\BibitemShut {NoStop}%
\bibitem [{\citenamefont {Guzzo}\ and\ \citenamefont {Lega}(2014)}]{Guzzo2014}%
  \BibitemOpen
  \bibfield  {author} {\bibinfo {author} {\bibfnamefont {M.}~\bibnamefont
  {Guzzo}}\ and\ \bibinfo {author} {\bibfnamefont {E.}~\bibnamefont {Lega}},\
  }\href {https://doi.org/10.1137/130930224} {\bibfield  {journal} {\bibinfo
  {journal} {SIAM Journal on Applied Mathematics}\ }\textbf {\bibinfo {volume}
  {74}},\ \bibinfo {pages} {1058} (\bibinfo {year} {2014})}\BibitemShut
  {NoStop}%
\bibitem [{\citenamefont {Demian}\ and\ \citenamefont
  {Wiggins}(2017)}]{Demian2017}%
  \BibitemOpen
  \bibfield  {author} {\bibinfo {author} {\bibfnamefont {A.~S.}\ \bibnamefont
  {Demian}}\ and\ \bibinfo {author} {\bibfnamefont {S.}~\bibnamefont
  {Wiggins}},\ }\href {https://doi.org/10.1142/S021812741750225X} {\bibfield
  {journal} {\bibinfo  {journal} {International Journal of Bifurcation and
  Chaos}\ }\textbf {\bibinfo {volume} {27}},\ \bibinfo {pages} {1750225}
  (\bibinfo {year} {2017})}\BibitemShut {NoStop}%
\bibitem [{\citenamefont {Lega}\ and\ \citenamefont {Guzzo}(2016)}]{Lega2017}%
  \BibitemOpen
  \bibfield  {author} {\bibinfo {author} {\bibfnamefont {E.}~\bibnamefont
  {Lega}}\ and\ \bibinfo {author} {\bibfnamefont {M.}~\bibnamefont {Guzzo}},\
  }\href@noop {} {\bibfield  {journal} {\bibinfo  {journal} {Proceedings of the
  XVII National Conference of Astronomers of Serbia}\ } (\bibinfo {year}
  {2016})}\BibitemShut {NoStop}%
\bibitem [{\citenamefont {Soley}\ and\ \citenamefont
  {Heller}(2018)}]{Soley2018}%
  \BibitemOpen
  \bibfield  {author} {\bibinfo {author} {\bibfnamefont {M.~B.}\ \bibnamefont
  {Soley}}\ and\ \bibinfo {author} {\bibfnamefont {E.~J.}\ \bibnamefont
  {Heller}},\ }\href {https://doi.org/10.1103/PhysRevA.98.052702} {\bibfield
  {journal} {\bibinfo  {journal} {Physical Review A}\ }\textbf {\bibinfo
  {volume} {98}},\ \bibinfo {pages} {052702} (\bibinfo {year}
  {2018})}\BibitemShut {NoStop}%
\bibitem [{\citenamefont {Soley}(2020)}]{Soley_thesis}%
  \BibitemOpen
  \bibfield  {author} {\bibinfo {author} {\bibfnamefont {M.~B.~S.}\
  \bibnamefont {Soley}},\ }\href@noop {} {\bibinfo {title} {Escaping from an
  ultracold inferno: Classical and semiclassical mechanics near threshold}}
  (\bibinfo {year} {2020})\BibitemShut {NoStop}%
\bibitem [{\citenamefont {Mancho}\ \emph {et~al.}(2013)\citenamefont {Mancho},
  \citenamefont {Wiggins}, \citenamefont {Curbelo},\ and\ \citenamefont
  {Mendoza}}]{mancho2013lagrangian}%
  \BibitemOpen
  \bibfield  {author} {\bibinfo {author} {\bibfnamefont {A.~M.}\ \bibnamefont
  {Mancho}}, \bibinfo {author} {\bibfnamefont {S.}~\bibnamefont {Wiggins}},
  \bibinfo {author} {\bibfnamefont {J.}~\bibnamefont {Curbelo}},\ and\ \bibinfo
  {author} {\bibfnamefont {C.}~\bibnamefont {Mendoza}},\ }\href
  {https://doi.org/10.1016/j.cnsns.2013.05.002} {\bibfield  {journal} {\bibinfo
   {journal} {Communications in Nonlinear Science and Numerical Simulation}\
  }\textbf {\bibinfo {volume} {18}},\ \bibinfo {pages} {3530} (\bibinfo {year}
  {2013})}\BibitemShut {NoStop}%
\bibitem [{\citenamefont {Agaoglou}\ \emph
  {et~al.}(2020{\natexlab{a}})\citenamefont {Agaoglou}, \citenamefont
  {Aguilar-Sanjuan}, \citenamefont {Garc{\'i}a-Garrido}, \citenamefont
  {Gonz{\'a}lez-Montoya}, \citenamefont {Katsanikas}, \citenamefont
  {Krajňák}, \citenamefont {Naik},\ and\ \citenamefont
  {Wiggins}}]{ldbook2020}%
  \BibitemOpen
  \bibfield  {author} {\bibinfo {author} {\bibfnamefont {M.}~\bibnamefont
  {Agaoglou}}, \bibinfo {author} {\bibfnamefont {B.}~\bibnamefont
  {Aguilar-Sanjuan}}, \bibinfo {author} {\bibfnamefont {V.~J.}\ \bibnamefont
  {Garc{\'i}a-Garrido}}, \bibinfo {author} {\bibfnamefont {F.}~\bibnamefont
  {Gonz{\'a}lez-Montoya}}, \bibinfo {author} {\bibfnamefont {M.}~\bibnamefont
  {Katsanikas}}, \bibinfo {author} {\bibfnamefont {V.}~\bibnamefont
  {Krajňák}}, \bibinfo {author} {\bibfnamefont {S.}~\bibnamefont {Naik}},\
  and\ \bibinfo {author} {\bibfnamefont {S.}~\bibnamefont {Wiggins}},\ }\href
  {https://doi.org/10.5281/zenodo.3958985} {\emph {\bibinfo {title} {Lagrangian
  Descriptors: Discovery and Quantification of Phase Space Structure and
  Transport}}}\ (\bibinfo  {publisher} {zenodo: 10.5281/zenodo.3958985},\
  \bibinfo {year} {2020})\BibitemShut {NoStop}%
\bibitem [{\citenamefont {Lopesino}\ \emph {et~al.}(2015)\citenamefont
  {Lopesino}, \citenamefont {Balibrea}, \citenamefont {Wiggins},\ and\
  \citenamefont {Mancho}}]{Lopesino2015}%
  \BibitemOpen
  \bibfield  {author} {\bibinfo {author} {\bibfnamefont {C.}~\bibnamefont
  {Lopesino}}, \bibinfo {author} {\bibfnamefont {F.}~\bibnamefont {Balibrea}},
  \bibinfo {author} {\bibfnamefont {S.}~\bibnamefont {Wiggins}},\ and\ \bibinfo
  {author} {\bibfnamefont {A.~M.}\ \bibnamefont {Mancho}},\ }\href
  {https://doi.org/10.1016/j.cnsns.2015.02.022} {\bibfield  {journal} {\bibinfo
   {journal} {Communications in Nonlinear Science and Numerical Simulation}\
  }\textbf {\bibinfo {volume} {27}},\ \bibinfo {pages} {40} (\bibinfo {year}
  {2015})}\BibitemShut {NoStop}%
\bibitem [{\citenamefont {Lopesino}\ \emph {et~al.}(2017)\citenamefont
  {Lopesino}, \citenamefont {Balibrea-Iniesta}, \citenamefont
  {Garc\'ia-Garrido}, \citenamefont {Wiggins},\ and\ \citenamefont
  {Mancho}}]{lopesino2017}%
  \BibitemOpen
  \bibfield  {author} {\bibinfo {author} {\bibfnamefont {C.}~\bibnamefont
  {Lopesino}}, \bibinfo {author} {\bibfnamefont {F.}~\bibnamefont
  {Balibrea-Iniesta}}, \bibinfo {author} {\bibfnamefont {V.~J.}\ \bibnamefont
  {Garc\'ia-Garrido}}, \bibinfo {author} {\bibfnamefont {S.}~\bibnamefont
  {Wiggins}},\ and\ \bibinfo {author} {\bibfnamefont {A.~M.}\ \bibnamefont
  {Mancho}},\ }\href {https://doi.org/10.1142/S0218127417300014} {\bibfield
  {journal} {\bibinfo  {journal} {International Journal of Bifurcation and
  Chaos}\ }\textbf {\bibinfo {volume} {27}},\ \bibinfo {pages} {1730001}
  (\bibinfo {year} {2017})}\BibitemShut {NoStop}%
\bibitem [{\citenamefont {Garc{\'{i}}a-Garrido}\ \emph
  {et~al.}(2018)\citenamefont {Garc{\'{i}}a-Garrido}, \citenamefont
  {Balibrea-Iniesta}, \citenamefont {Wiggins}, \citenamefont {Mancho},\ and\
  \citenamefont {Lopesino}}]{Garcia-Garrido2018}%
  \BibitemOpen
  \bibfield  {author} {\bibinfo {author} {\bibfnamefont {V.~J.}\ \bibnamefont
  {Garc{\'{i}}a-Garrido}}, \bibinfo {author} {\bibfnamefont {F.}~\bibnamefont
  {Balibrea-Iniesta}}, \bibinfo {author} {\bibfnamefont {S.}~\bibnamefont
  {Wiggins}}, \bibinfo {author} {\bibfnamefont {A.~M.}\ \bibnamefont
  {Mancho}},\ and\ \bibinfo {author} {\bibfnamefont {C.}~\bibnamefont
  {Lopesino}},\ }\href {https://doi.org/10.1134/S1560354718060096} {\bibfield
  {journal} {\bibinfo  {journal} {Regular and Chaotic Dynamics}\ }\textbf
  {\bibinfo {volume} {23}},\ \bibinfo {pages} {751} (\bibinfo {year}
  {2018})}\BibitemShut {NoStop}%
\bibitem [{\citenamefont {Balibrea-Iniesta}\ \emph {et~al.}(2016)\citenamefont
  {Balibrea-Iniesta}, \citenamefont {Lopesino}, \citenamefont {Wiggins},\ and\
  \citenamefont {Mancho}}]{Iniesta2016}%
  \BibitemOpen
  \bibfield  {author} {\bibinfo {author} {\bibfnamefont {F.}~\bibnamefont
  {Balibrea-Iniesta}}, \bibinfo {author} {\bibfnamefont {C.}~\bibnamefont
  {Lopesino}}, \bibinfo {author} {\bibfnamefont {S.}~\bibnamefont {Wiggins}},\
  and\ \bibinfo {author} {\bibfnamefont {A.~M.}\ \bibnamefont {Mancho}},\
  }\href {https://doi.org/10.1142/S0218127416300366} {\bibfield  {journal}
  {\bibinfo  {journal} {International Journal of Bifurcation and Chaos}\
  }\textbf {\bibinfo {volume} {26}},\ \bibinfo {pages} {1630036} (\bibinfo
  {year} {2016})}\BibitemShut {NoStop}%
\bibitem [{\citenamefont {Mendoza}\ and\ \citenamefont
  {Mancho}(2010)}]{mendoza2010}%
  \BibitemOpen
  \bibfield  {author} {\bibinfo {author} {\bibfnamefont {C.}~\bibnamefont
  {Mendoza}}\ and\ \bibinfo {author} {\bibfnamefont {A.~M.}\ \bibnamefont
  {Mancho}},\ }\href {https://doi.org/10.1103/PhysRevLett.105.038501}
  {\bibfield  {journal} {\bibinfo  {journal} {Phys. Rev. Lett.}\ }\textbf
  {\bibinfo {volume} {105}},\ \bibinfo {pages} {038501} (\bibinfo {year}
  {2010})}\BibitemShut {NoStop}%
\bibitem [{\citenamefont {Naik}\ \emph {et~al.}(2019)\citenamefont {Naik},
  \citenamefont {Garc\'ia-Garrido},\ and\ \citenamefont {Wiggins}}]{naik2019a}%
  \BibitemOpen
  \bibfield  {author} {\bibinfo {author} {\bibfnamefont {S.}~\bibnamefont
  {Naik}}, \bibinfo {author} {\bibfnamefont {V.~J.}\ \bibnamefont
  {Garc\'ia-Garrido}},\ and\ \bibinfo {author} {\bibfnamefont {S.}~\bibnamefont
  {Wiggins}},\ }\href {https://doi.org/10.1016/j.cnsns.2019.104907} {\bibfield
  {journal} {\bibinfo  {journal} {Communications in Nonlinear Science and
  Numerical Simulation}\ }\textbf {\bibinfo {volume} {79}},\ \bibinfo {pages}
  {104907} (\bibinfo {year} {2019})}\BibitemShut {NoStop}%
\bibitem [{\citenamefont {Katsanikas}\ \emph
  {et~al.}(2020{\natexlab{a}})\citenamefont {Katsanikas}, \citenamefont
  {Garc\'{i}a-Garrido},\ and\ \citenamefont {Wiggins}}]{katsanikas2020a}%
  \BibitemOpen
  \bibfield  {author} {\bibinfo {author} {\bibfnamefont {M.}~\bibnamefont
  {Katsanikas}}, \bibinfo {author} {\bibfnamefont {V.~J.}\ \bibnamefont
  {Garc\'{i}a-Garrido}},\ and\ \bibinfo {author} {\bibfnamefont
  {S.}~\bibnamefont {Wiggins}},\ }\href
  {https://doi.org/https://doi.org/10.1016/j.cplett.2020.137199} {\bibfield
  {journal} {\bibinfo  {journal} {Chemical Physics Letters}\ }\textbf {\bibinfo
  {volume} {743}},\ \bibinfo {pages} {137199} (\bibinfo {year}
  {2020}{\natexlab{a}})}\BibitemShut {NoStop}%
\bibitem [{\citenamefont {Katsanikas}\ \emph
  {et~al.}(2020{\natexlab{b}})\citenamefont {Katsanikas}, \citenamefont
  {Garc{\'\i}a-Garrido},\ and\ \citenamefont
  {Wiggins}}]{katsanikas2020detection}%
  \BibitemOpen
  \bibfield  {author} {\bibinfo {author} {\bibfnamefont {M.}~\bibnamefont
  {Katsanikas}}, \bibinfo {author} {\bibfnamefont {V.~J.}\ \bibnamefont
  {Garc{\'\i}a-Garrido}},\ and\ \bibinfo {author} {\bibfnamefont
  {S.}~\bibnamefont {Wiggins}},\ }\href
  {https://doi.org/10.1142/S0218127420300268} {\bibfield  {journal} {\bibinfo
  {journal} {International Journal of Bifurcation and Chaos}\ }\textbf
  {\bibinfo {volume} {30}},\ \bibinfo {pages} {2030026} (\bibinfo {year}
  {2020}{\natexlab{b}})}\BibitemShut {NoStop}%
\bibitem [{\citenamefont {Katsanikas}\ \emph
  {et~al.}(2020{\natexlab{c}})\citenamefont {Katsanikas}, \citenamefont
  {Garc{\'\i}a-Garrido}, \citenamefont {Agaoglou},\ and\ \citenamefont
  {Wiggins}}]{katsanikas2020phase}%
  \BibitemOpen
  \bibfield  {author} {\bibinfo {author} {\bibfnamefont {M.}~\bibnamefont
  {Katsanikas}}, \bibinfo {author} {\bibfnamefont {V.~J.}\ \bibnamefont
  {Garc{\'\i}a-Garrido}}, \bibinfo {author} {\bibfnamefont {M.}~\bibnamefont
  {Agaoglou}},\ and\ \bibinfo {author} {\bibfnamefont {S.}~\bibnamefont
  {Wiggins}},\ }\href {https://doi.org/10.1103/PhysRevE.102.012215} {\bibfield
  {journal} {\bibinfo  {journal} {Physical Review E}\ }\textbf {\bibinfo
  {volume} {102}},\ \bibinfo {pages} {012215} (\bibinfo {year}
  {2020}{\natexlab{c}})}\BibitemShut {NoStop}%
\bibitem [{\citenamefont {Crossley}\ \emph {et~al.}(2021)\citenamefont
  {Crossley}, \citenamefont {Agaoglou}, \citenamefont {Katsanikas},\ and\
  \citenamefont {Wiggins}}]{crossley2021poincare}%
  \BibitemOpen
  \bibfield  {author} {\bibinfo {author} {\bibfnamefont {R.}~\bibnamefont
  {Crossley}}, \bibinfo {author} {\bibfnamefont {M.}~\bibnamefont {Agaoglou}},
  \bibinfo {author} {\bibfnamefont {M.}~\bibnamefont {Katsanikas}},\ and\
  \bibinfo {author} {\bibfnamefont {S.}~\bibnamefont {Wiggins}},\ }\href
  {https://doi.org/10.1134/S1560354721020040} {\bibfield  {journal} {\bibinfo
  {journal} {Regular and Chaotic Dynamics}\ }\textbf {\bibinfo {volume} {26}},\
  \bibinfo {pages} {147} (\bibinfo {year} {2021})}\BibitemShut {NoStop}%
\bibitem [{\citenamefont {Naik}\ and\ \citenamefont
  {Wiggins}(2020)}]{naik2020}%
  \BibitemOpen
  \bibfield  {author} {\bibinfo {author} {\bibfnamefont {S.}~\bibnamefont
  {Naik}}\ and\ \bibinfo {author} {\bibfnamefont {S.}~\bibnamefont {Wiggins}},\
  }\href {https://doi.org/10.1039/D0CP01362E} {\bibfield  {journal} {\bibinfo
  {journal} {Phys. Chem. Chem. Phys.}\ }\textbf {\bibinfo {volume} {22}},\
  \bibinfo {pages} {17890} (\bibinfo {year} {2020})}\BibitemShut {NoStop}%
\bibitem [{\citenamefont {Krajňák}\ \emph {et~al.}(2019)\citenamefont
  {Krajňák}, \citenamefont {Ezra},\ and\ \citenamefont
  {Wiggins}}]{krajnak2019}%
  \BibitemOpen
  \bibfield  {author} {\bibinfo {author} {\bibfnamefont {V.}~\bibnamefont
  {Krajňák}}, \bibinfo {author} {\bibfnamefont {G.}~\bibnamefont {Ezra}},\
  and\ \bibinfo {author} {\bibfnamefont {S.}~\bibnamefont {Wiggins}},\ }\href
  {https://doi.org/https://doi.org/10.1134/S1560354719060030} {\bibfield
  {journal} {\bibinfo  {journal} {Regul. Chaot. Dyn.}\ }\textbf {\bibinfo
  {volume} {24}},\ \bibinfo {pages} {615–627} (\bibinfo {year}
  {2019})}\BibitemShut {NoStop}%
\bibitem [{\citenamefont {Garc\'{i}a-Garrido}\ \emph
  {et~al.}(2020{\natexlab{a}})\citenamefont {Garc\'{i}a-Garrido}, \citenamefont
  {Naik},\ and\ \citenamefont {Wiggins}}]{GG2020a}%
  \BibitemOpen
  \bibfield  {author} {\bibinfo {author} {\bibfnamefont {V.~J.}\ \bibnamefont
  {Garc\'{i}a-Garrido}}, \bibinfo {author} {\bibfnamefont {S.}~\bibnamefont
  {Naik}},\ and\ \bibinfo {author} {\bibfnamefont {S.}~\bibnamefont
  {Wiggins}},\ }\href {https://doi.org/10.1142/S0218127420300086} {\bibfield
  {journal} {\bibinfo  {journal} {International Journal of Bifurcation and
  Chaos}\ }\textbf {\bibinfo {volume} {30}},\ \bibinfo {pages} {2030008}
  (\bibinfo {year} {2020}{\natexlab{a}})}\BibitemShut {NoStop}%
\bibitem [{\citenamefont {Garc\'{i}a-Garrido}\ \emph
  {et~al.}(2020{\natexlab{b}})\citenamefont {Garc\'{i}a-Garrido}, \citenamefont
  {Agaoglou},\ and\ \citenamefont {Wiggins}}]{GG2020b}%
  \BibitemOpen
  \bibfield  {author} {\bibinfo {author} {\bibfnamefont {V.~J.}\ \bibnamefont
  {Garc\'{i}a-Garrido}}, \bibinfo {author} {\bibfnamefont {M.}~\bibnamefont
  {Agaoglou}},\ and\ \bibinfo {author} {\bibfnamefont {S.}~\bibnamefont
  {Wiggins}},\ }\href
  {https://doi.org/https://doi.org/10.1016/j.cnsns.2020.105331} {\bibfield
  {journal} {\bibinfo  {journal} {Communications in Nonlinear Science and
  Numerical Simulation}\ }\textbf {\bibinfo {volume} {89}},\ \bibinfo {pages}
  {105331} (\bibinfo {year} {2020}{\natexlab{b}})}\BibitemShut {NoStop}%
\bibitem [{\citenamefont {Agaoglou}\ \emph
  {et~al.}(2020{\natexlab{b}})\citenamefont {Agaoglou}, \citenamefont
  {Garc{\'\i}a-Garrido}, \citenamefont {Katsanikas},\ and\ \citenamefont
  {Wiggins}}]{agaoglou2020phase}%
  \BibitemOpen
  \bibfield  {author} {\bibinfo {author} {\bibfnamefont {M.}~\bibnamefont
  {Agaoglou}}, \bibinfo {author} {\bibfnamefont {V.}~\bibnamefont
  {Garc{\'\i}a-Garrido}}, \bibinfo {author} {\bibfnamefont {M.}~\bibnamefont
  {Katsanikas}},\ and\ \bibinfo {author} {\bibfnamefont {S.}~\bibnamefont
  {Wiggins}},\ }\href
  {https://doi.org/https://doi.org/10.1016/j.cplett.2020.137610} {\bibfield
  {journal} {\bibinfo  {journal} {Chemical Physics Letters}\ }\textbf {\bibinfo
  {volume} {754}},\ \bibinfo {pages} {137610} (\bibinfo {year}
  {2020}{\natexlab{b}})}\BibitemShut {NoStop}%
\bibitem [{\citenamefont {Agaoglou}\ \emph {et~al.}(2019)\citenamefont
  {Agaoglou}, \citenamefont {Aguilar-Sanjuan}, \citenamefont
  {Garc{\'i}a-Garrido}, \citenamefont {Garc{\'i}a-Meseguer}, \citenamefont
  {Gonz{\'a}lez-Montoya}, \citenamefont {Katsanikas}, \citenamefont
  {Krajňák}, \citenamefont {Naik},\ and\ \citenamefont
  {Wiggins}}]{agaoglou2019}%
  \BibitemOpen
  \bibfield  {author} {\bibinfo {author} {\bibfnamefont {M.}~\bibnamefont
  {Agaoglou}}, \bibinfo {author} {\bibfnamefont {B.}~\bibnamefont
  {Aguilar-Sanjuan}}, \bibinfo {author} {\bibfnamefont {V.~J.}\ \bibnamefont
  {Garc{\'i}a-Garrido}}, \bibinfo {author} {\bibfnamefont {R.}~\bibnamefont
  {Garc{\'i}a-Meseguer}}, \bibinfo {author} {\bibfnamefont {F.}~\bibnamefont
  {Gonz{\'a}lez-Montoya}}, \bibinfo {author} {\bibfnamefont {M.}~\bibnamefont
  {Katsanikas}}, \bibinfo {author} {\bibfnamefont {V.}~\bibnamefont
  {Krajňák}}, \bibinfo {author} {\bibfnamefont {S.}~\bibnamefont {Naik}},\
  and\ \bibinfo {author} {\bibfnamefont {S.}~\bibnamefont {Wiggins}},\ }\href
  {https://doi.org/10.5281/zenodo.3568210} {\emph {\bibinfo {title} {Chemical
  Reactions: A Journey into Phase Space}}}\ (\bibinfo  {publisher} {zenodo:
  10.5281/zenodo.3568210},\ \bibinfo {year} {2019})\BibitemShut {NoStop}%
\bibitem [{\citenamefont {Montoya}\ and\ \citenamefont
  {Wiggins}(2020)}]{Montoya2020}%
  \BibitemOpen
  \bibfield  {author} {\bibinfo {author} {\bibfnamefont {F.~G.}\ \bibnamefont
  {Montoya}}\ and\ \bibinfo {author} {\bibfnamefont {S.}~\bibnamefont
  {Wiggins}},\ }\href {https://doi.org/10.1103/PhysRevE.102.062203} {\bibfield
  {journal} {\bibinfo  {journal} {Physical Review E}\ }\textbf {\bibinfo
  {volume} {102}},\ \bibinfo {pages} {62203} (\bibinfo {year}
  {2020})}\BibitemShut {NoStop}%
\bibitem [{\citenamefont {{Gonzalez Montoya}}\ and\ \citenamefont
  {Wiggins}(2020)}]{GonzalezMontoya2020}%
  \BibitemOpen
  \bibfield  {author} {\bibinfo {author} {\bibfnamefont {F.}~\bibnamefont
  {{Gonzalez Montoya}}}\ and\ \bibinfo {author} {\bibfnamefont
  {S.}~\bibnamefont {Wiggins}},\ }\bibfield  {journal} {\bibinfo  {journal}
  {Journal of Physics A: Mathematical and Theoretical}\ }\href
  {https://doi.org/10.1088/1751-8121/ab8b75} {10.1088/1751-8121/ab8b75}
  (\bibinfo {year} {2020})\BibitemShut {NoStop}%
\bibitem [{LDD()}]{LDDS}%
  \BibitemOpen
  \href@noop {} {\bibinfo {title} {{LDDS}: {P}ython package for computing and
  visualizing {L}agrangian {D}escriptors in {D}ynamical {S}ystems.}},\ \bibinfo
  {howpublished} {\url{https://github.com/champsproject/ldds}}\BibitemShut
  {NoStop}%
\bibitem [{sci()}]{scipy_laplace}%
  \BibitemOpen
  \href@noop {} {}\bibinfo {howpublished}
  {\url{https://docs.scipy.org/doc/scipy/reference/generated/scipy.ndimage.laplace.html}}\BibitemShut
  {NoStop}%
\bibitem [{\citenamefont {Skokos}\ \emph {et~al.}(2002)\citenamefont {Skokos},
  \citenamefont {Patsis},\ and\ \citenamefont
  {Athanassoula}}]{skokos2002orbital}%
  \BibitemOpen
  \bibfield  {author} {\bibinfo {author} {\bibfnamefont {C.}~\bibnamefont
  {Skokos}}, \bibinfo {author} {\bibfnamefont {P.}~\bibnamefont {Patsis}},\
  and\ \bibinfo {author} {\bibfnamefont {E.}~\bibnamefont {Athanassoula}},\
  }\href@noop {} {\bibfield  {journal} {\bibinfo  {journal} {Monthly Notices of
  the Royal Astronomical Society}\ }\textbf {\bibinfo {volume} {333}},\
  \bibinfo {pages} {847} (\bibinfo {year} {2002})}\BibitemShut {NoStop}%
\bibitem [{\citenamefont {Contopoulos}(2004)}]{contopoulos2004order}%
  \BibitemOpen
  \bibfield  {author} {\bibinfo {author} {\bibfnamefont {G.}~\bibnamefont
  {Contopoulos}},\ }\href@noop {} {\emph {\bibinfo {title} {Order and chaos in
  dynamical astronomy}}}\ (\bibinfo  {publisher} {Springer Science \& Business
  Media},\ \bibinfo {year} {2004})\BibitemShut {NoStop}%
\bibitem [{\citenamefont {Katsanikas}\ and\ \citenamefont
  {Wiggins}(2018)}]{katsanikas2018phase}%
  \BibitemOpen
  \bibfield  {author} {\bibinfo {author} {\bibfnamefont {M.}~\bibnamefont
  {Katsanikas}}\ and\ \bibinfo {author} {\bibfnamefont {S.}~\bibnamefont
  {Wiggins}},\ }\href {https://doi.org/10.1142/S0218127418300422} {\bibfield
  {journal} {\bibinfo  {journal} {International Journal of Bifurcation and
  Chaos}\ }\textbf {\bibinfo {volume} {28}},\ \bibinfo {pages} {1830042}
  (\bibinfo {year} {2018})}\BibitemShut {NoStop}%
\bibitem [{\citenamefont {Kolmogorov}(1954)}]{kolmogorov1954}%
  \BibitemOpen
  \bibfield  {author} {\bibinfo {author} {\bibfnamefont {A.}~\bibnamefont
  {Kolmogorov}},\ }\href@noop {} {\bibfield  {journal} {\bibinfo  {journal}
  {Dokl. Akad. Nauk SSSR}\ }\textbf {\bibinfo {volume} {98}},\ \bibinfo {pages}
  {527} (\bibinfo {year} {1954})}\BibitemShut {NoStop}%
\bibitem [{\citenamefont {Arnold}(1963)}]{arnold1963}%
  \BibitemOpen
  \bibfield  {author} {\bibinfo {author} {\bibfnamefont {V.~I.}\ \bibnamefont
  {Arnold}},\ }\href@noop {} {\bibfield  {journal} {\bibinfo  {journal}
  {Russian Mathematical Surveys}\ }\textbf {\bibinfo {volume} {18}},\ \bibinfo
  {pages} {9} (\bibinfo {year} {1963})}\BibitemShut {NoStop}%
\bibitem [{\citenamefont {Moser}(1962)}]{moser1962}%
  \BibitemOpen
  \bibfield  {author} {\bibinfo {author} {\bibfnamefont {J.~K.}\ \bibnamefont
  {Moser}},\ }\href@noop {} {\bibfield  {journal} {\bibinfo  {journal} {Nachr.
  Akad. Wiss. G\"{o}ttingen II, Math. Phys. Kl.}\ ,\ \bibinfo {pages} {1}}
  (\bibinfo {year} {1962})}\BibitemShut {NoStop}%
\bibitem [{\citenamefont {Liapounoff}(1907)}]{liapounoff1907probleme}%
  \BibitemOpen
  \bibfield  {author} {\bibinfo {author} {\bibfnamefont {A.}~\bibnamefont
  {Liapounoff}},\ }in\ \href@noop {} {\emph {\bibinfo {booktitle} {Annales de
  la Facult{\'e} des sciences de Toulouse: Math{\'e}matiques}}},\ Vol.~\bibinfo
  {volume} {9}\ (\bibinfo {year} {1907})\ pp.\ \bibinfo {pages}
  {203--474}\BibitemShut {NoStop}%
\bibitem [{\citenamefont {Moser}(1958)}]{moser1958generalization}%
  \BibitemOpen
  \bibfield  {author} {\bibinfo {author} {\bibfnamefont {J.}~\bibnamefont
  {Moser}},\ }\href {https://doi.org/10.1002/cpa.3160110208} {\bibfield
  {journal} {\bibinfo  {journal} {Communications on Pure and Applied
  Mathematics}\ }\textbf {\bibinfo {volume} {11}},\ \bibinfo {pages} {257}
  (\bibinfo {year} {1958})}\BibitemShut {NoStop}%
\bibitem [{\citenamefont {Kelley}(1967)}]{kelley1967liapounov}%
  \BibitemOpen
  \bibfield  {author} {\bibinfo {author} {\bibfnamefont {A.}~\bibnamefont
  {Kelley}},\ }\href@noop {} {\bibfield  {journal} {\bibinfo  {journal}
  {Journal of mathematical analysis and applications}\ }\textbf {\bibinfo
  {volume} {18}},\ \bibinfo {pages} {472} (\bibinfo {year} {1967})}\BibitemShut
  {NoStop}%
\end{thebibliography}%



\end{document}